\documentclass[aps,prx,reprint,superscriptaddress,longbibliography]{revtex4-2}
\usepackage{amsmath,amssymb,graphicx, bm,xcolor,hyperref}
\usepackage[utf8]{inputenc}
\usepackage[T1]{fontenc}
\usepackage{braket}
\usepackage{booktabs}
\usepackage[normalem]{ulem}
\usepackage{siunitx}

\begin{document}

\title{Scalable Fluxonium Quantum Processors via Tunable-Coupler Architecture}
\author{Ze Zhan}
\thanks{These authors contributed equally to this work.}
\affiliation{Quantum Science Center of Guangdong-Hong Kong-Macao Greater Bay Area, Shenzhen, China}

\author{Zishuo Li}
\thanks{These authors contributed equally to this work.}
\affiliation{Quantum Science Center of Guangdong-Hong Kong-Macao Greater Bay Area, Shenzhen, China}

\author{Fei Wang}
\author{Wangwei Lan}
\author{Xianchuang Pan}
\author{Liang Xiang}
\author{Xu Dou}
\author{Ran Gao}
\author{Guicheng Gong}
\author{Yanbo Guo}
\author{Quan Guan}
\author{Lijuan Hu}
\author{Ruizhi Hu}
\author{Honghong Ji}
\author{Lijing Jin}
\author{Yongyue Jin}
\author{Chengyao Li}
\author{Kannan Lu}
\author{Lu Ma}
\author{Xizheng Ma}
\author{Hongcheng Wang}
\author{Jiahui Wang}
\author{Huijuan Zhan}
\author{Tao Zhou}
\author{Xing Zhu}

\author{Chunqing Deng}
\email{dengchunqing@quantumsc.cn}
\affiliation{Quantum Science Center of Guangdong-Hong Kong-Macao Greater Bay Area, Shenzhen, China}

\author{Tenghui Wang}
\email{wangtenghui@quantumsc.cn}
\affiliation{Quantum Science Center of Guangdong-Hong Kong-Macao Greater Bay Area, Shenzhen, China}

\date{\today}

\begin{abstract}
Superconducting quantum processors have largely converged on transmon-based architectures, while alternative qubit modalities with intrinsic error protection have lacked a demonstrated path to scalable system integration. In particular, although tunable-coupler–mediated interactions have been validated for small fluxonium systems, it remains unclear whether such designs can be scaled to a multi-qubit lattice. 
Here, we establish a scalable fluxonium processor architecture based on a modular qubit–coupler unit cell engineered to suppress residual interactions and spectator errors in a many-qubit lattice. The system enables parallel single-qubit gate fidelities approaching 99.99\% and two-qubit CZ gate fidelities around 99\%. With an optimized gate duration of 32~ns, the best CZ gate fidelity reaches 99.9\%. We further validate this architecture in a 22-qubit processor based on the same configuration, where parallel operations enable the deterministic generation of Greenberger–Horne–Zeilinger states involving up to 10 qubits. Together, these results demonstrate that the fluxonium–tunable-coupler unit cell composes without emergent interaction pathologies and establish fluxonium as a scalable superconducting qubit platform.
\end{abstract}

\maketitle

\section{Introduction}

Superconducting circuits~\cite{Kjaergaard2020} constitute one of the leading platforms for the realization of quantum processors, owing to their strong interactions that enable fast gate operations, the flexibility of circuit design for system-level engineering, and compatibility with large-scale lithographic fabrication. Over the past decade, most scalable superconducting quantum processors have converged on architectures based on the transmon qubit~\cite{Koch2007}—a simple circuit consisting of a Josephson junction shunted by a capacitance. This architectural convergence has enabled rapid progress in processor scaling, high-fidelity parallel quantum operations~\cite{arute2019quantum, Wu2021}, and demonstrations of increasingly complex quantum algorithms~\cite{kim2023evidence, Abanin2025, Jin2025} and error-correcting protocols~\cite{Acharya2025, He2025, Besedin2026, Wang2026}. However, as these systems scale, limitations associated with the transmon's weak anharmonicity and dense capacitive coupling—manifested as spectral crowding and unexpected coupling —begin to constrain parallel operation fidelity and raise risks about architectural scalability.

Fluxonium qubits~\cite{manucharyan2009coherent} have attracted considerable interest as an alternative superconducting qubit due to their noise protection properties, large anharmonicity, and rich energy spectra. These features have enabled demonstrations of long coherence times~\cite{Somoroff2023, Wang2025} and high-fidelity control in two-qubit systems~\cite{Ficheux2021, bao2022fluxonium, moskalenko2022high, ding2023high, ma2023native, zhang2024tunable, Lin2025}. However, despite these advantages, the development of scalable fluxonium-based processors has remained limited, in part due to the challenge of engineering controllable multi-qubit interactions while preserving qubit isolation. Recently, tunable-coupler–mediated interactions have been demonstrated between fluxonium qubits, enabling high-fidelity entangling gates together with strong suppression of residual coupling at the level of a few qubits~\cite{moskalenko2022high, ding2023high, zhang2024tunable}. Nevertheless, it remains unclear whether such designs can be composed into larger multi-qubit lattices without emergent interaction pathologies.

In this work we establish a scalable fluxonium processor architecture based on a modular fluxonium–transmon–fluxonium (FTF) unit cell. The architecture integrates fluxonium qubits with tunable couplers to enable controllable interactions while preserving qubit isolation. Leveraging the rich multilevel structure of the fluxonium spectrum, we implement a spectral allocation strategy that simultaneously supports high-fidelity qubit operations, including gates, reset, and measurement. In addition, flux-pulse–activated couplers provide a large on–off interaction ratio, enabling fast entangling gates while maintaining strong suppression of residual coupling during idle operation in multi-qubit arrays. 

We first validate this architecture in a four-qubit processor, demonstrating average parallel single-qubit gate fidelities around 99.99\% and two-qubit CZ gate fidelities around 99\%, with a maximum fidelity of 99.9\% at a gate duration of 32~ns. We then extend the same unit-cell design to a 22-qubit processor, where we verify that both residual ZZ couplings and spectator-induced errors remain strongly suppressed across the extended system. Parallel operations further enable the deterministic generation of Greenberger--Horne--Zeilinger (GHZ) states involving up to 10 qubits. These results demonstrate that the fluxonium--tunable-coupler unit cell composes into extended multi-qubit lattices without introducing emergent interaction pathologies, establishing fluxonium as a scalable superconducting qubit platform.

\begin{figure*}[bt]
    \includegraphics[width = \textwidth]{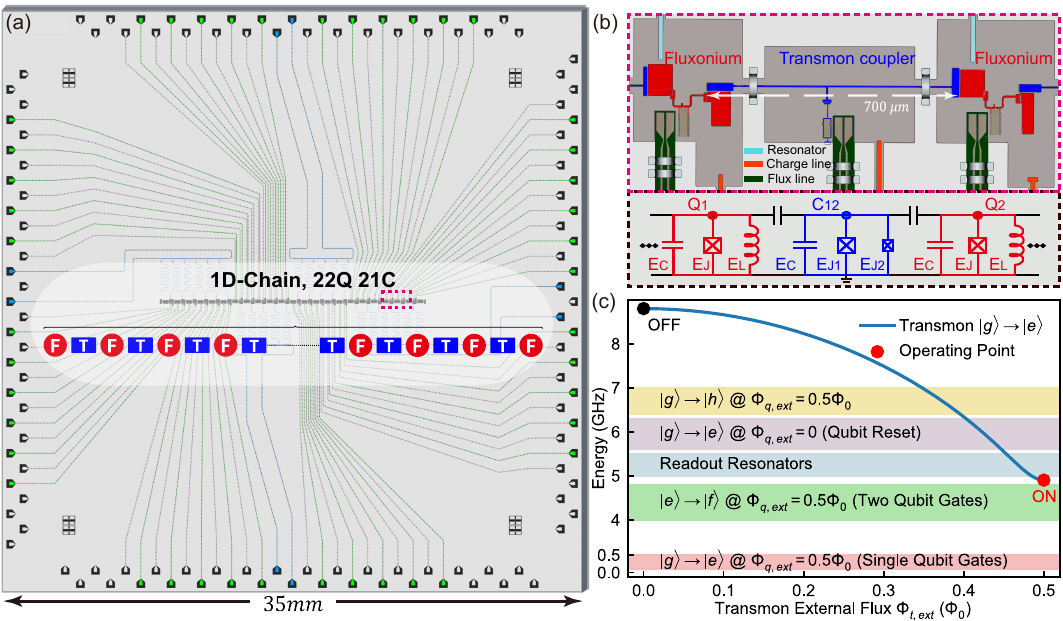}
    \caption{\textbf{Processor architecture.} (a) Device layout of the 22-qubit fluxonium processor comprising 21 transmon couplers arranged in a one-dimensional chain. For notational clarity, qubits are indexed from $Q_1$ to $Q_{22}$ in a left-to-right order, and each coupler $C_{ij}$ is denoted by the indices of the neighboring qubits it connects.  Each repeating unit consists of two fluxonium qubits capacitively coupled via an intermediate transmon coupler, forming a modular architecture that can be extended to larger systems. 
    (b) Upper panel: Optical micrograph of the corresponding device. The coupler and qubit pads, flux and charge lines, and readout resonators are highlighted using false colors. The maximum span of the coupler is approximately $700~\mu m$, as indicated by the white arrow. Lower panel: Circuit diagram of a single fluxonium–transmon–fluxonium (FTF) unit. Each fluxonium qubit consists of a Josephson junction shunted by a capacitance and a large linear inductance, characterized by Josephson energy $E_J$, charging energy $E_C$, and inductive energy $E_L$, respectively. The two qubits are capacitively coupled to a tunable transmon coupler formed by two asymmetric Josephson junctions with energies $E_{J1}$ and $E_{J2}$.
    (c) Frequency allocation and operating regime of the processor. The fluxonium $\ket{g}\!\leftrightarrow\!\ket{e}$ transitions lie in the 100--500\,MHz range (red), while the higher excited-state transitions $\ket{e}\!\leftrightarrow\!\ket{f}$ and $\ket{g}\!\leftrightarrow\!\ket{h}$ occur in the 4--5\,GHz (green) and 6.3--7.0\,GHz (yellow) ranges, respectively. Readout resonators are centered near 5–5.5\,GHz (blue). The transmon coupler provides a tunable frequency band, with its maximum exceeding 8\,GHz and its minimum positioned between the qubit $\ket{e}\!\leftrightarrow\!\ket{f}$ transitions. In the coupling-OFF configuration, the coupler is biased above 8\,GHz, where it is far detuned from the qubit transitions, thereby suppressing residual qubit–qubit interactions. During gate operation (coupling-ON configuration), the coupler is brought close to the $\ket{e}\!\leftrightarrow\!\ket{f}$ transitions (green) by tuning its flux bias $\Phi_{t,\mathrm{ext}}$ to approximately $0.5~\Phi_0$. Fluxonium qubit frequencies at zero flux (purple) are set above the readout resonators and can be flux-tuned into resonance for fast state initialization.}
    \label{fig1}
\end{figure*}

\section{Device Architecture}

To implement scalable, high-fidelity multi-qubit operations, we design a 22-qubit processor incorporating 21 tunable couplers (Fig.~\ref{fig1}(a)). The processor is built from repeating FTF units arranged in a one-dimensional chain. Neighboring fluxonium qubits are coupled capacitively through a transmon coupler, which acts as the only interaction channel between them. Fig.~\ref{fig1}(b) shows a scalable FTF unit, including an optical micrograph (upper panel) and the corresponding circuit diagram (lower panel). Each qubit and coupler has an independent flux-bias line for frequency tuning and a dedicated charge-drive line for microwave drive.

Distinct from previous implementations \cite{ding2023high}, our design employs an increased qubit–qubit separation, i.e., $\sim700~\mu$m, which significantly suppresses direct capacitive coupling between qubits. This strategy is particularly well suited to fluxonium qubits, whose intrinsically limited capacitance budget renders direct coupling pathways negligible and thus calls for an alternative “turn-off” mechanism.
In contrast to conventional transmon-based systems \cite{arute2019quantum, Wu2021,yan2018tunable} —where effective decoupling typically arises from destructive interference between direct capacitive coupling and indirect, coupler-mediated interactions—we operate the coupler in a regime of large detuning from the fluxonium plasmon mode. In this limit, the effective interaction is parametrically suppressed, enabling efficient decoupling without relying on interference between competing pathways. This detuning-based mechanism provides a robust route to controllable coupling in FTF architectures.

The operational principle of this architecture is illustrated by the spectral structure in Fig.~\ref{fig1}(c). When biased at half-integer flux quanta, fluxonium computational states correspond to low-frequency fluxon transitions (100--500~MHz, red)~\cite{manucharyan2009coherent}. The coupler, having a higher eigenfrequency, couples only weakly to these modes. As a result, neighboring fluxonium qubits remain effectively isolated, suppressing residual ZZ and XX interactions and enabling high-fidelity parallel single-qubit gates. In contrast, the higher-frequency plasmon transitions of fluxonium (4--5~GHz, green) can be brought into resonance with the coupler, forming a hybridized subsystem that mediates controlled two-qubit interactions. The transmon coupler features junction asymmetry, giving rise to a tunable frequency band whose maximum exceeds 8~GHz in the coupling-OFF configuration and whose minimum allows tuning close to the plasmon transitions in the coupling-ON configuration. 

In particular, the qubit–coupler interaction is exploited using a microwave-activated CZ scheme, in which a resonant drive couples selected computational states to a hybridized manifold involving noncomputational (plasmon) states of the qubit and excitations of the coupler. This selective excitation induces a controlled, state-dependent phase accumulation while remaining off-resonant for other states, thereby suppressing unwanted transitions and leakage. For state characterization and reset, readout resonators are centered near 5--5.5~GHz (blue). Fluxonium $\ket{g}\leftrightarrow\ket{e}$ frequencies at zero flux are set above the readout resonators (purple) and can be flux-tuned into resonance for fast state initialization~\cite{Reed2010fast,mcewen2021removing}. This allocation scheme exploits the fluxonium energy hierarchy to interleave operational frequencies across well-separated spectral bands, enabling decoupled calibration, mitigating parameter interdependence, and enhancing robustness against fabrication variations.

To ensure the feasibility of the described spectral allocation, we specify representative Hamiltonian parameters for both the fluxonium qubits and the transmon couplers. The fluxonium qubits are characterized by inductive energy $E_L/h \approx 0.55$--$0.7$~GHz, charging energy $E_C/h \approx 1.0$--$1.3$~GHz, and Josephson energy $E_J/h \approx 4.0$--$5.0$~GHz. The transmon couplers are defined by Josephson energies $E_{J1}/h\approx 16.5$--$17$~GHz and $E_{J2}/h\approx 28$--$28.5$~GHz and a charging energy $E_C/h\approx 0.21$--$0.23$~GHz. A detailed numerical analysis together with a specific parameter set is provided in the Appendix \ref{Spectral}.

\section{Parallel Operations in a Modular Chip}

\begin{figure*}
  \includegraphics[width = \textwidth]{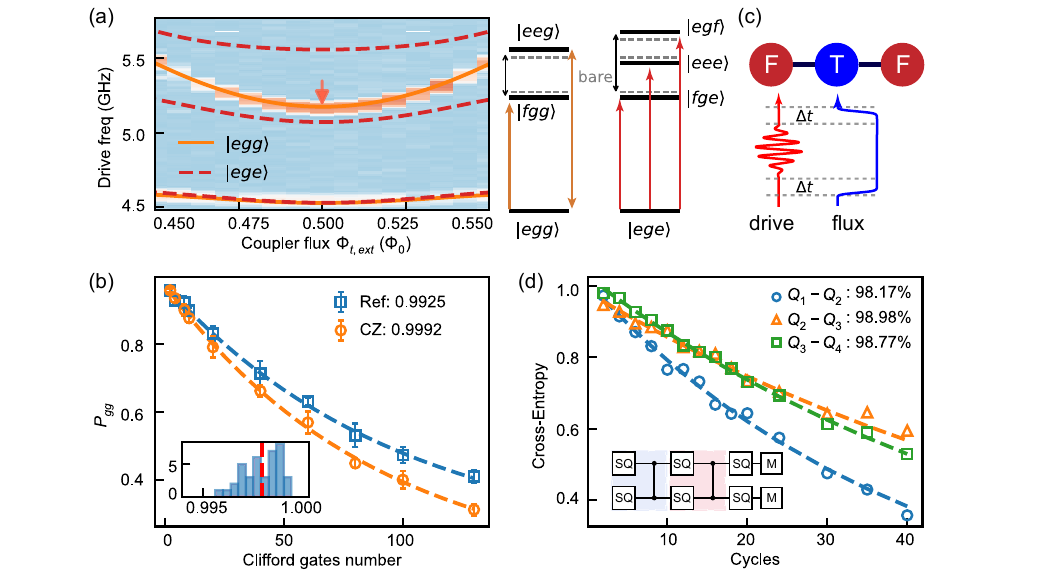}
  \caption{\textbf{High-fidelity and parallel two-qubit gate operation} (a) Flux-dependent transition spectrum of the relevant noncomputational states in the $Q_2$–$C_{23}$–$Q_3$ subsystem. The color map shows measured transition strengths for excitations initialized from $\ket{egg}$. The two dominant spectral branches correspond to transitions to $\ket{fgg}$ (lower) and $\ket{eeg}$ (upper), indicated by solid lines. At the coupling-ON operating point, hybridization between the $\ket{egf}$ (upper dashed line) and $\ket{eee}$ (middle dashed line) states gives rise to a state-dependent transition, enabling the CZ gate (marked by the arrow).
  (b) Interleaved randomized benchmarking of the CZ gate at the operating point. The gate achieves a best fidelity exceeding 99.9\% with a duration of 32 ns, demonstrating high-fidelity two-qubit control within the fluxonium–coupler architecture. The inset shows repeated measurements, yielding an average fidelity of 99.78\% (red dashed line).
  (c) Experimental control scheme for the parallel CZ gate. A square flux-bias pulse brings the transmon coupler into resonance with the fluxonium $\ket{e}\leftrightarrow\ \ket{f}$ transition, activating hybridization between the qubit and coupler states. Microwave drives applied to the coupler and fluxonium then selectively excite state-dependent transitions, resulting in a CZ gate.
  (d) Random circuit sampling (RCS) for parallel gate benchmarking. Each cycle consists of a layer of random single-qubit gates followed by a CZ gate (inset). Parallel RCS is performed on the $Q_1$–$Q_2$ and $Q_3$–$Q_4$ pairs, while individual RCS is performed on $Q_2$–$Q_3$. The extracted error-per-cycle fidelities are $98.16\%$, $98.78\%$, and $98.98\%$, respectively. A buffer time $\Delta t=10$~ns is inserted between flux and drive pulse for each CZ gate, and the microwave drive duration is 60~ns.} \label{fig2}
\end{figure*}

We first validate high-fidelity CZ gates and parallel operations in a four-qubit modular processor arranged as a one-dimensional fluxonium–transmon–fluxonium chain, representing a subsystem of the 22-qubit architecture. This four-qubit chain constitutes a representative subarray of the full processor, incorporating the same qubit–coupler unit cells, control wiring, and spectral allocation strategy, and thus captures the essential architectural features required for scalable operation.
Fig.~\ref{fig2}(a) shows the flux-dependent transition spectrum of the $Q_2$–$C_{23}$–$Q_3$ subsystem (see Appendix \ref{4Q}). Starting from the $\ket{egg}$ state ($\ket{Q_2C_{23}Q_3}$), two dominant transitions are observed as the coupler flux is varied, corresponding to single-photon excitations into the dressed $\ket{eeg}$ and $\ket{fgg}$ states. The observed level repulsion arises from hybridization between the fluxonium plasmon transition $\ket{e}\!\leftrightarrow\!\ket{f}$ and the coupler excitation, consistent with theoretical modeling (solid curves).

The transition structure is further clarified by comparing excitations originating from the $\ket{egg}$ and $\ket{ege}$ states (Fig.~\ref{fig2}(a) Right). Starting from $\ket{egg}$, the two observed hybridized branches correspond to transitions to $\ket{{fgg}}$ and $\ket{{eeg}}$. Starting from $\ket{ege}$, three transitions to $\ket{fge}$, $\ket{eee}$, and $\ket{egf}$ emerge. The splittings of the $\ket{egg}\!\leftrightarrow\!\ket{eeg}$ and $\ket{ege}\!\leftrightarrow\!\ket{eee}$ transitions are different because the $\ket{eee}$ state hybridizes simultaneously with $\ket{fge}$ and $\ket{egf}$, which correspond to the plasmon transitions of the neighboring fluxonium qubits and the coupler excitation. These branches are well separated when the coupler is biased near $\Phi_{t,\mathrm{ext}} = 0.5\,\Phi_0$, corresponding to the coupling-ON configuration, enabling identification of an isolated operating point. At the selected operating point (arrow in Fig.~\ref{fig2}(a)), resonant driving selectively couples $\ket{egg}$ to $\ket{{eeg}}$, resulting in coherent oscillations within the hybridized subspace. A controlled $2\pi$ rotation in this subspace imparts a conditional $\pi$ phase on $\ket{egg}$, thereby implementing a microwave-activated CZ gate. 
To benchmark the intrinsic fidelity limit, we first fix the coupler $C_{23}$ at $\Phi_{t,\mathrm{ext}} = 0.5\,\Phi_0$. With an optimized gate duration of 32~ns, the CZ fidelity measured from interleaved randomized benchmarking exceeds 99.9\% at its best (Fig.~\ref{fig2}(b))~\cite{magesan2012efficient, barends2014surface}. Repeated measurements yield an average fidelity of 99.78\% (red dashed line), with the distribution shown in the inset.

For parallel gate operations, the couplers are kept in the coupling-OFF configuration during single-qubit and idle periods, where they are biased above 8~GHz to suppress residual interactions. During CZ gate execution, the couplers are selectively activated via flux pulses. A square flux pulse (blue in Fig.~\ref{fig2}(c)) brings the coupler to the coupling-ON configuration for a duration of 80~ns, while a cosine-shaped microwave pulse of 60~ns (red in Fig.~\ref{fig2}(c)) drives transitions to the hybridized states during the flat-top region, thereby implementing the CZ gate.

We then perform random circuit sampling (RCS) on two qubit pairs simultaneously, as illustrated in the inset of Fig.~\ref{fig2}(d). Each cycle consists of a layer of random single-qubit (SQ) gates followed by a CZ gate, with a final random rotation applied before joint readout. The SQ gates are decomposed into two $X_{\pi/2}$ rotations, one $X_{\pi}$ rotation, and virtual-$Z$ gates~\cite{Chen2023Compiling}. Independent characterization of four qubits under simultaneous driving yields single-qubit fidelities at the level of $99.99\%$ (see Appendix \ref{4Q}). The extracted error-per-cycle fidelities are $98.16\%$ for the $Q_1$–$Q_2$ pair and $98.78\%$ for the $Q_3$–$Q_4$ pair, with corresponding purity estimates of $98.11\%$ and $98.81\%$, indicating operation close to the decoherence limit. For consistency, we also benchmark the $Q_2$–$Q_3$ pair individually, obtaining an RCS fidelity of $98.98\%$, comparable to the parallel results. These results demonstrate strong parallel operability in the modular architecture, where CZ interactions can be selectively activated in a controlled and on-demand manner without compromising concurrent operations.

\section{Suppression of Residual Interactions and Spectator Errors in a Multi-Qubit Lattice}

\begin{figure}
  \includegraphics[width = 0.48\textwidth]{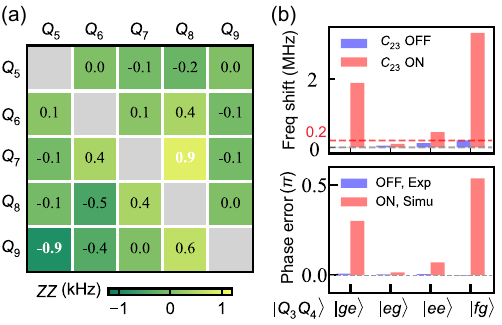}
  \caption{ \textbf{Suppression of residual interactions and spectator errors in a multi-qubit lattice.} (a) Measured residual ZZ interaction matrix in a five-qubit chain ($Q_5$–$Q_9$). The horizontal (vertical) axis labels the control (target) qubit. (b) Spectator-induced frequency shifts and controlled-phase errors during a CZ gate between $Q_1$–$Q_2$ as a function of the spectator state of $Q_3$–$Q_4$. The intermediate coupler $C_{23}$ is either in the coupling-OFF or coupling-ON configuration. Frequency shifts are defined relative to the reference case with spectator qubits initialized in $\ket{gg}$, and controlled-phase errors are defined as deviations from the ideal $\pi$ phase for the CZ gate. In the coupling-ON configuration, the controlled-phase error is not directly measured but extracted from the induced frequency shifts using simulations assuming a CZ gate duration of 120~ns.}
  \label{fig3}
\end{figure}

To evaluate residual interactions and spectator-induced errors which are key limitations for scalable parallel operation, we characterize both residual ZZ and XX couplings, as well as dynamic crosstalk during two-qubit operations in the 22-qubit processor. We randomly select two representative qubit chains from the 22-qubit device, $Q_5$--$Q_9$ and $Q_1$--$Q_4$, to characterize residual ZZ crosstalk (Fig.~\ref{fig3}(a)) and spectator-induced crosstalk (Fig.~\ref{fig3}(b)), respectively. We first quantify residual interactions by measuring residual ZZ couplings in the $Q_5$--$Q_9$ chain. The ZZ interaction strength is extracted using a Ramsey-type measurement of the target qubit frequency shift conditioned on the state of a neighboring qubit. As shown in Fig.~\ref{fig3}(a), all measured ZZ couplings are below $\sim1~\mathrm{kHz}$  when the couplers are biased in the coupling-OFF configuration. We further investigate the residual ZZ coupling when the couplers are tuned to the coupling-ON configuration. Similar to the coupling-OFF configuration, the residual ZZ coupling remains strongly suppressed at the $\sim1~\mathrm{kHz}$ level when the coupler is biased in the coupling-ON configuration, as detailed in the Appendix \ref{Spectator Errors}. Given their small magnitude, the extracted values are partially limited by the coherence time of the target qubit, leading to minor asymmetries in the measured interaction matrix. 

We also probe residual XX-type coupling by tuning neighboring qubits into resonance and find it to be below 10~kHz (see Appendix \ref{Couplings}). Consistent with these results, simultaneous single-qubit gate fidelities reach the 99.9\% level (see Appendix \ref{22Q}). The observed suppression of residual interactions within the computational subspace arises from the combination of capacitive coupling and the small charge dipole matrix elements associated with the low-frequency fluxonium qubit transitions. Moreover, the large anharmonicity of fluxonium energetically separates the computational states from the higher excited states that are primarily responsible for residual ZZ, leading to strong suppression of both unwanted ZZ and exchange-type couplings.

We next examine the isolation of two-qubit gate operations by directly quantifying spectator-induced errors from neighboring qubits during CZ gate execution. Experimentally, we monitor the transition frequency required to implement $Q_1$-$Q_2$ controlled phase gate as well as the value of the controlled phase while preparing the spectator qubits $Q_3$-$Q_4$ in different states. During this measurement, the relevant couplers $C_{12}$ and $C_{34}$ are activated. Fig.~\ref{fig3}(b) shows the resulting transition frequency shifts and the controlled phase error relative to the reference case with spectators initialized in $\ket{gg}$.

When $C_{23}$ is in the coupling-OFF configuration, both the transition frequency shift and the resulting controlled-phase error remain negligible across all spectator states. The spectator-induced frequency shifts are suppressed below 200~kHz, with the corresponding phase error remaining below 0.005~rad, indicating that identical control parameters produce a consistent CZ gate. In contrast, when $C_{23}$ is activated, the transition frequency exhibits a strong dependence on the spectator state, preventing a single parameter set from implementing a consistent CZ gate. To quantify this effect, we incorporate the experimentally measured frequency shifts as detunings in numerical simulations (see Appendix~\ref{Spectator Errors}). Using a pulse duration of 120~ns, we extract the deviation of the accumulated controlled phase from $\pi$ for a complete cyclic trajectory. Both the frequency shifts and the inferred phase errors provide direct measures of spectator-induced crosstalk. Together, these results demonstrate that the coupler suppresses spectator-induced errors in the coupling-OFF configuration, which is essential for scalable parallel quantum computation.
The suppression of spectator errors arises from tuning the coupler to effectively decouple the fluxonium plasmon transitions that mediate the two-qubit interaction, analogous to the cancellation mechanism employed in tunable-coupler transmon architectures.

\section{Generation of Multi-Qubit GHZ States}

\begin{figure*}
  \includegraphics[width = \textwidth]{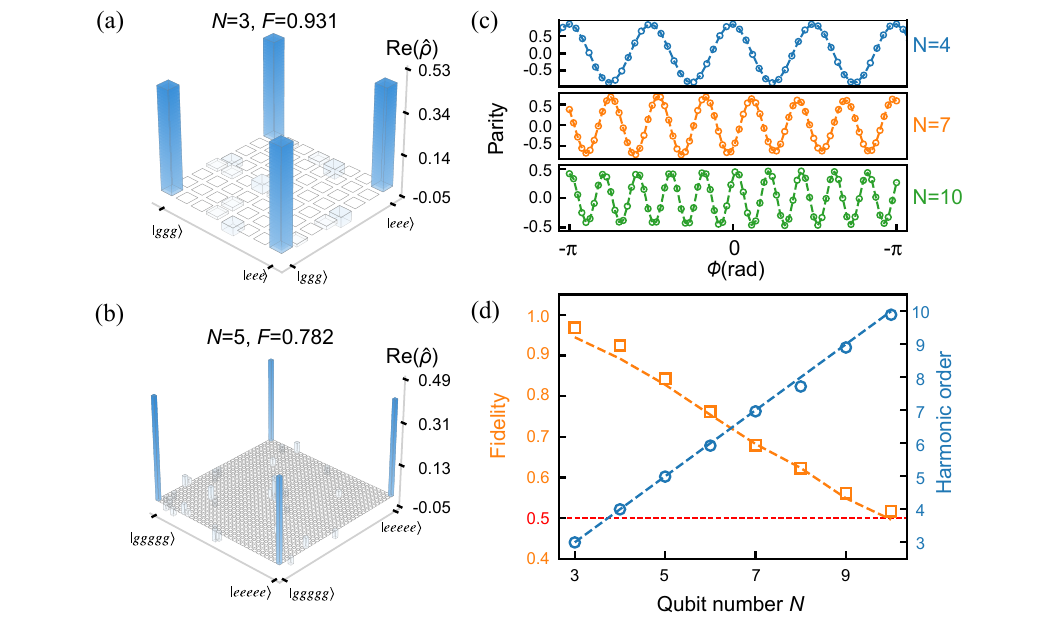}
  \caption{\textbf{Generation and characterization of multi-qubit GHZ states.} (a), (b) Reconstructed real part of the density matrix $\hat{\rho}$ for the prepared 3- and 5-qubit GHZ states obtained via quantum state tomography. 
  (c) Measured parity as a function of the applied collective phase $\phi$ for GHZ states with qubit numbers $N = 4,\, 7,\,$ and $10$. (d) Extracted harmonic order and GHZ-state fidelity as a function of qubit number $N$. The harmonic order scales linearly with $N$, while the fidelity decreases with system size due to accumulated errors. Dashed lines correspond to numerical simulations.}
  \label{fig4}
\end{figure*}

We next perform a system-level validation of controllability by realizing $N$-qubit entanglement. GHZ states provide a stringent benchmark, as their preparation requires coordinated initialization, entangling gates, and readout across all participating qubits. In the experiment, a single qubit is initialized in $(\ket{g} + \ket{e})/\sqrt{2}$, and sequential CNOT gates between adjacent qubits (implemented via CZ gates with Hadamard rotations on the target) generate an $N$-qubit GHZ state along the chain. The detailed circuit and calibration procedures are provided in the Appendix~\ref{GHZ}.

For small system sizes ($N \leqslant 5$), we perform quantum state tomography to verify genuine multipartite entanglement~\cite{dicarlo2010preparation, barends2014surface}. The reconstructed density matrices $\hat\rho$ for the 3- and 5-qubit GHZ states are shown in Fig.~\ref{fig4}(a) and (b), exhibiting the expected GHZ structure with dominant populations and coherences between $\ket{gg\cdots g}$ and $\ket{ee\cdots e}$.
For larger systems, where tomography becomes impractical, we probe the coherence of the generated states using parity measurements. In this protocol, a collective phase rotation is applied to the GHZ state $R_Z^N=\bigotimes_{i=1}^{N} R_z^{(i)}(\phi)$, preparing a final state $\hat\rho_f$. This is followed by a measurement of the parity operator $\Pi_N = \bigotimes_{i = 1} ^N R_X^{(i)}(\pi/2)$. For an $N$-qubit GHZ state, the parity signal exhibits coherent oscillations of the form
\begin{equation}
\mathrm{Tr} [\hat\rho_f(\phi) \Pi_N ] = A_N\cos(N\phi),
\end{equation}
where the oscillation frequency scales linearly with the number of entangled qubits, providing a direct signature of global phase coherence, and the oscillation amplitude $A_N$ quantifies the multi-qubit coherence of the state. In the current device, GHZ states are demonstrated up to $N=10$ qubits along the chain $Q_{12}$--$Q_{21}$.

In Fig.~\ref{fig4}(c), we measure the parity as a function of the applied collective phase $\phi$ for GHZ states with different qubit numbers. The data show clear oscillations whose frequency increases with $N$, while the amplitude $A_N$ decreases with system size due to accumulated errors. Combining $A_N$ with the measured populations $P_{g,N} = \rho_{gg\dots g,gg\dots g}$ and $P_{e,N} = \rho_{ee\dots e,ee\dots e}$, we estimate the GHZ-state fidelity as $F_N = (P_{g,N} + P_{e,N} + A_N)/2$~\cite{song201710,wei2020verifying}. Fig.~\ref{fig4}(d) compares the measured fidelities with an error model (orange dashed line) based on independently characterized single- and two-qubit gate fidelities (see Appendix~\ref{GHZ}), taking into account the total gate count for each $N$. The good agreement indicates that the fidelity decay is primarily limited by the average CZ gate performance, with minimal additional contributions from parallel operation overhead or residual inter-qubit coupling. For $N=10$, the measured fidelity of $52\%$ exceeds the threshold for genuine multipartite entanglement ($F > 0.5$). This represents the largest demonstration of multi-qubit entanglement in a fluxonium-based processor and extends the scale of coherent multipartite operations in this platform.

\section{Conclusion and outlook}

We have demonstrated a scalable fluxonium-based superconducting quantum processor architecture built upon a modular fluxonium–tunable-coupler unit cell. The architecture enables strong and controllable two-qubit interactions while preserving effective isolation of qubits outside gate operations. We experimentally validate this design across multiple scales, from few-qubit subsystems to a 22-qubit processor. High-fidelity parallel operations are achieved together with uniformly suppressed residual interactions and negligible spectator-induced errors in the coupling-OFF regime. The generation of a 10-qubit GHZ state further confirms coherent multi-qubit control at the system level.

The observed advantages stem from a distinct operating regime intrinsic to our fluxonium-based architecture. Unlike transmon architectures that often require fine-tuned interaction cancellation and complex multi-qubit calibration overhead~\cite{Klimov2024, valles2025}, our design intrinsically suppresses residual ZZ and unwanted exchange interactions. This is achieved through the combination of capacitive coupling, small charge dipole matrix elements of low-frequency transitions, and the large anharmonicity of fluxonium qubits. Consequently, qubit and coupler frequencies can be arranged in a simple, scalable manner without emergent interaction pathologies. Taken together, these results establish that the proposed unit cell composes into extended multi-qubit systems without interaction-induced degradation, providing a highly viable pathway toward scalable fluxonium quantum processors.

Looking forward, extending this architecture to two-dimensional layouts is a key step toward scalable quantum processors. Such scaling requires increased qubit spacing to accommodate control wiring while maintaining strong and controllable coupling between multiple neighbors. This, in turn, necessitates suppressing stray capacitance to preserve the intended coupling structure. These challenges can be addressed by employing a double-transmon coupler design that enables larger qubit spacing while maintaining strong nearest-neighbor interactions, as demonstrated in Ref.~\cite{GXC2026}. Together, these advances provide a viable pathway toward two-dimensional fluxonium-based processors with scalable connectivity and control.

\section{Data Availability Statement}
The data that support the findings of this study are available from the corresponding author upon reasonable request.

\section*{Acknowledgments}
The authors thank Ziang Wang, Feng Wu, Tian Xia, and Hui-Hai Zhao for their helpful discussions and valuable suggestions regarding the chip design. This research was supported by the Guangdong Provincial Quantum Science Strategic Initiative (Grant No. GDZX2407001). The authors also express their gratitude to the Westlake Center for Micro/Nano Fabrication, Zhejiang QizhenTek Co., Ltd., and the Micro-Nano Fabrication and Device Center at the Songshan Lake Materials Laboratory for their essential technical support during the chip fabrication process. We thank Talent Microwave Inc.\ for their essential support in microwave components and Yunmao Technology Co.\ for providing fabrication tools. We thank Dr. Yaoyun Shi for his participation in the planning of the project, and his contribution to securing the funding.

\section*{Author Contributions}
C. Deng and T. Wang supervised the project. T. Wang conducted the experiments on the 4-qubit modular chip. Z. Zhan and Z. Li calibrated the 22-qubit processor and performed the experiments on GHZ state generation. X. Pan, L. Xiang, J. Wang, and R. Hu carried out measurements and system characterization. L. Jin, W. Lan, Q. Guan, X. Dou, and X. Ma designed the chip. F. Wang, L. Ma, Y. Jin, H. Zhan, and C. Li fabricated the chip. R. Gao, H. Ji, and L. Hu performed the chip packaging. X. Zhu, T. Zhou, G. Gong, H. Wang, and K. Lu developed the control electronics. C. Deng, T. Wang, Z. Zhan, Z. Li, and L. Jin wrote the manuscript with input from all authors.

\appendix


\title{Supplementary Material for \\``Scalable Fluxonium Quantum Processors via Tunable-Coupler Architecture"}

\maketitle

\section{Spectral Selectivity and Spectator-Error Suppression in the FTF Unit} \label{Spectral}

\begin{figure}
  \includegraphics[width = 0.48\textwidth]{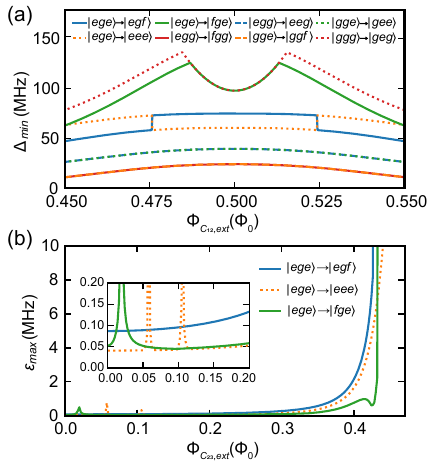} \caption{\textbf{Spectral properties of MAP gate transitions and spectator-error in the FTF unit.}
    (a) Minimum spectral detuning $\Delta_{\mathrm{min}}$ between the target two-qubit transitions and the nearest competing parasitic transitions as a function of the external magnetic flux bias $\Phi_{C_{12}, \mathrm{ext}}$ of the active coupler $C_{12}$. The solid, dashed, and dotted lines correspond to different initial states and their corresponding relevant transitions, demonstrating sufficient spectral separation around the operating point ($\Phi_{C_{12}, \mathrm{ext}}/\Phi_0 \approx 0.5$) for high-fidelity, selective microwave activation. 
    (b) Maximum spectator-induced frequency shift $\epsilon_{\mathrm{max}}$ on the active qubit pair ($Q_1, Q_2$) arising from the state of the spectator qubit $Q_3$, plotted as a function of the external flux bias $\Phi_{C_{23}, \mathrm{ext}}$ applied to the intermediate coupler $C_{23}$. The inset highlights the range $\Phi_{C23,\mathrm{ext}} = 0.0$–$0.2,\Phi_0$, corresponding to the coupling-OFF configuration, where the spectator-induced shifts are suppressed to the $\sim 100$~kHz level, indicating effective isolation of the MAP transitions during parallel operation.}\label{FigS0}
\end{figure}

This section analyzes the spectral properties of the fluxonium–transmon–fluxonium (FTF) unit relevant to the microwave-activated phase (MAP) gate scheme. We focus on two key aspects: (i) spectral selectivity of the driven transition and (ii) suppression of spectator-induced frequency shifts in the coupling-OFF configuration. 

To realize a high-fidelity CZ gate, a target transition must be selectively driven while suppressing leakage to nearby transitions. This selectivity is determined by two factors: the minimum spectral separation from competing transitions, and the corresponding transition matrix elements. To quantify this effect, we consider a two-qubit system $(Q_1, Q_2)$ coupled via the transmon coupler $C_{12}$. We characterize the spectral selectivity by the minimum detuning
\begin{equation}
\Delta_{\rm min} = \min_{i'\rightarrow f' \neq i\rightarrow f} \left| E_{i'\rightarrow f'} - E_{i \rightarrow f} \right| / h,
\end{equation}
where $E_{i \rightarrow f}$ denotes the transition energy between eigenstates adiabatically connected to the corresponding bare product states. Fig.~\ref{FigS0}(a) shows $\Delta_{\rm min}$ as a function of the coupler external flux bias. When the external flux is tuned to the coupling-ON regime (near half a flux quantum), the target transition is well separated from nearby unintended transitions, with minimum detunings on the order of $\sim 100$~MHz. This spectral separation suppresses off-resonant excitation and enables selective activation of the desired transition, thereby minimizing leakage.

We next investigate the impact of spectator qubits on driven transitions. In multi-qubit architectures, residual interactions manifest as spectator-dependent frequency shifts, which limit the fidelity of parallel gate operations. To quantify this effect, we consider a three-qubit system $(Q_1, Q_2, Q_3)$ coupled via transmon couplers, where a CZ gate is applied to the active pair $(Q_1, Q_2)$ by driving the transition between the states $|A\rangle$ and $|B\rangle$ (see Fig.~\ref{FigS0}(b)), while $Q_3$ acts as a spectator. We define the maximum spectator-induced frequency shift as
\begin{equation}
\epsilon_{\rm max}
= \max_{\substack{i,j \in {g,e,f}, \ i \neq j}}
\left|
(E_{Bgj} - E_{Agj}) - (E_{Bgi} - E_{Agi})
\right|/h,
\end{equation}
where $E_{\Lambda g k}$ denotes the eigenenergy of the full system state adiabatically connected to the bare state
$|\Lambda\rangle_{\mathrm{active}} \otimes |g\rangle_{C_{23}} \otimes |k\rangle_{Q_3}$.
Here, $|\Lambda\rangle_{\mathrm{active}} \in {|A\rangle, |B\rangle}$ labels the driven subspace of the active qubit pair $(Q_1, Q_2)$, the coupler $C_{23}$ is restricted to its ground state, and $k \in {i,j}$ with $i,j \in {g,e,f}$ specifies the state of the spectator qubit $Q_3$. As shown in Fig.~\ref{FigS0}(b), $\epsilon_{\mathrm{max}}$ is strongly suppressed when the coupler is biased in the coupling-OFF configuration, reaching values on the order of $\sim 100$~kHz. This demonstrates that the MAP transitions are effectively insensitive to the spectator state, establishing a well-defined decoupled regime. Together, these results show that the chosen parameter regime simultaneously enables (i) well-resolved noncomputational levels for MAP gate and (ii) strong suppression of spectator-induced crosstalk, both of which are essential for high-fidelity parallel two-qubit operations.

The numerical simulations presented above are based on a representative set of Hamiltonian parameters consistent with the design principles outlined in the main text. The parameters are not obtained through fine-tuning, but instead reflect a physically realizable regime of the device. In particular, the chosen values naturally reproduce the key spectral features required for operation, including well-resolved hybridized levels for MAP gate transitions and strong suppression of spectator-induced frequency shifts in the coupling-OFF configuration. The full set of parameters used in the simulation is summarized in Table~\ref{tab:simu}.
\begin{table}[htbp]
\centering
\caption{Parameters of unit design}
\label{tab:simu}
\begin{tabular}{@{}lccccc@{}}
\toprule
Qubit & $Q_1$ & $C_{12}$ & $Q_2$ & $C_{23}$ & $Q_3$  \\
\midrule
$E_{C}$ ($h\cdot$\si{\giga\hertz}) & 1.17 & 0.21 & 1.3 & 0.23 & 1.3  \\
$E_{J}\backslash E_{J1}$ ($h\cdot$\si{\giga\hertz}) & 4.2 & 17 & 4 & 16.5 & 4.9 \\
$E_{L}\backslash E_{J2}$ ($h\cdot$\si{\giga\hertz}) & 0.6 & 28 & 0.55 & 28.5  & 0.55 \\
\midrule
Couplings ($h\cdot$\si{\giga\hertz}) & $Q_1$ & $C_{12}$ & $Q_2$ & $C_{23}$ & $Q_3$  \\
\midrule
$Q_1$ & $\backslash$ & 0.309 &-0.068 & 0& 0 \\
$C_{12}$ & 0.309 & $\backslash$ &-0.371 & 0.019& 0 \\
$Q_2$ & -0.068 & -0.371& $\backslash$ &  0.375& -0.065 \\
$C_{23}$ & 0 & 0.019 &0.375 & $\backslash$& -0.375\\
$Q_3$ & 0 & 0 &-0.065 & -0.375& $\backslash$ \\
\bottomrule
\end{tabular}
\end{table}

\section{Experimental Setup}
\label{Setup}

\begin{figure}
  \includegraphics[width = 0.48\textwidth]{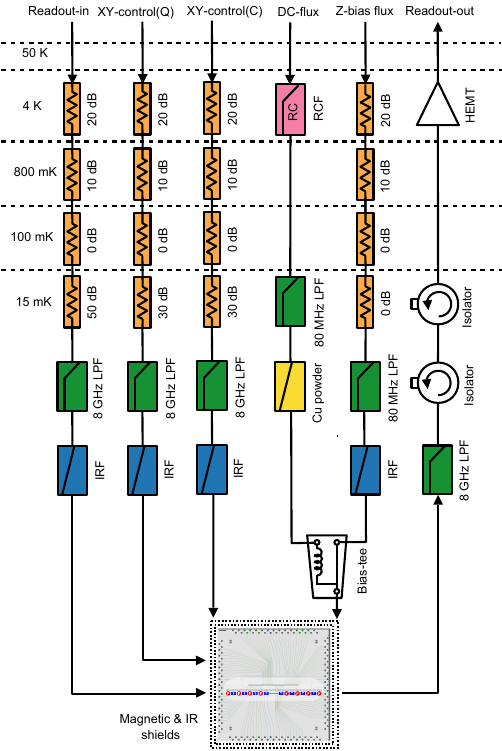} \caption{\label{experiment_setup}\textbf{Cryogenic setup.} Schematic detailing the low-temperature wiring inside the dilution refrigerator for operating the 22-qubit processor.}
\end{figure}

Experiments were performed using a setup as illustrated in Fig.~\ref{experiment_setup}. The 22-qubit device is anchored to the mixing chamber of a dilution refrigerator with a base temperature of $\sim 15\,\text{mK}$ and is shielded by a $\mu$-metal container. Qubit control is implemented using single-port arbitrary waveform generators (AWGs) to generate baseband $XY$-control and $Z$-bias flux tuning ($0\text{--}500\,\text{MHz}$), while microwave drives ($4\text{--}6\,\text{GHz}$) for readout and CZ gate operations are synthesized via IQ upconversion. 
Dedicated DC-flux lines provide current signals to bias the qubits or couplers at fixed flux points. All AWG ports and DC sources are provided by our home-made electronics. 
All input lines are attenuated and filtered across multiple temperature stages to suppress thermal noise. Specifically, the $XY$ control lines (for both qubits and couplers) incorporate $60\,\text{dB}$ of total attenuation, while the fast flux and readout input lines feature $30\,\text{dB}$ and $80\,\text{dB}$ attenuation, respectively. Additional broadband attenuation and thermalization are provided by a combination of RC filters, low-pass filters (LPF), copper-powder filters (Cu powder), and infrared filters (IRF). Output signals are routed through low-pass filters and circulators, amplified by a cryogenic high-electron-mobility transistor (HEMT) amplifier, and recorded using a high-speed digitizer.

\section{Fabrication}

The chip was fabricated on a 2-inch $c$-cut sapphire substrate. A $180~\mathrm{nm}$-thick tantalum film was sputtered onto the substrate, followed by spin-coating with $1~\mu\mathrm{m}$ of S1813 photoresist. All circuits except for the Josephson junctions were patterned via direct laser writing, and the development was carried out in MIF-319 developer, followed by a rinse with deionized (DI) water. Subsequently, the tantalum film was etched in a reactive ion etching (RIE) system using an Ar/BCl$_3$ gas mixture to form the base layer circuit. 

Al/AlO$_x$/Al Josephson junctions were fabricated using Manhattan-type shadow evaporation. The wafer was spin-coated with a bilayer electron-beam (e-beam) resist composed of PMMA A7 and MAA EL9, which was then patterned using a JOEL JBX8100FS e-beam lithography system. Development was performed by immersing the wafer in a 1:3 mixture of methyl isobutyl ketone (MIBK) and isopropyl alcohol (IPA) for 2 minutes, followed by a 2-minute IPA rinse. The patterned wafer was then loaded into an ultra-high vacuum (UHV) cluster deposition system for junction formation. A short ion milling step was conducted to remove resist residues and native oxide, ensuring galvanic contact between the Josephson junctions and capacitor pads. A $60~\mathrm{nm}$-thick Al base electrode layer was e-beam evaporated onto the wafer, followed by room-temperature (RT) oxidation at $6~\mathrm{Torr}$ for 40 minutes to form the barrier layer. Next, a $120~\mathrm{nm}$-thick Al counter electrode layer was e-beam evaporated, with a subsequent 20-minute RT oxidation at $20~\mathrm{Torr}$. The lift-off process was performed by sequentially immersing the wafer in baths of N-methyl-2-pyrrolidone (NMP), acetone, and IPA.

Following the junction fabrication, aluminum airbridges were constructed to realize low-loss crossovers in the superconducting circuitry. A first photolithography step defined the airbridge scaffold using SPR220-3 photoresist, spin-coated at 2300 rpm for $60~\mathrm{s}$ and soft-baked at $115~^\circ\mathrm{C}$ for $180~\mathrm{s}$. Patterns were exposed by direct-write lithography and developed in MIF-319 for $150~\mathrm{s}$. The resist was then thermally reflowed at $150~^\circ\mathrm{C}$ for $420~\mathrm{s}$ to form a rounded profile, promoting conformal Al deposition. Prior to metal deposition, the exposed surface was cleaned \textit{in situ} using ion beam etching to remove native oxides and contaminants. A $500~\mathrm{nm}$-thick Al layer was subsequently deposited by electron-beam evaporation to form the airbridge structural layer. A second photolithography step defined the lateral geometry of the airbridges, where SPR220-3 was spin-coated at 5000 rpm for $60~\mathrm{s}$ and soft-baked at $115~^\circ\mathrm{C}$ for $120~\mathrm{s}$. This was followed by exposure under the same conditions as the first layer and development in MIF-319 for $120~\mathrm{s}$. The Al film was patterned by wet etching using a Type A Al etchant for approximately $10~\mathrm{min}$ to remove unprotected regions. Finally, the resist scaffold was stripped via oxygen plasma ashing followed by solvent cleaning, thereby releasing the suspended airbridge structures.

\section{Single- and Two-Qubit Gate Performance in Four-Qubit Modular chip}
\label{4Q}

\begin{figure*}
  \includegraphics[width = 1.0\textwidth]{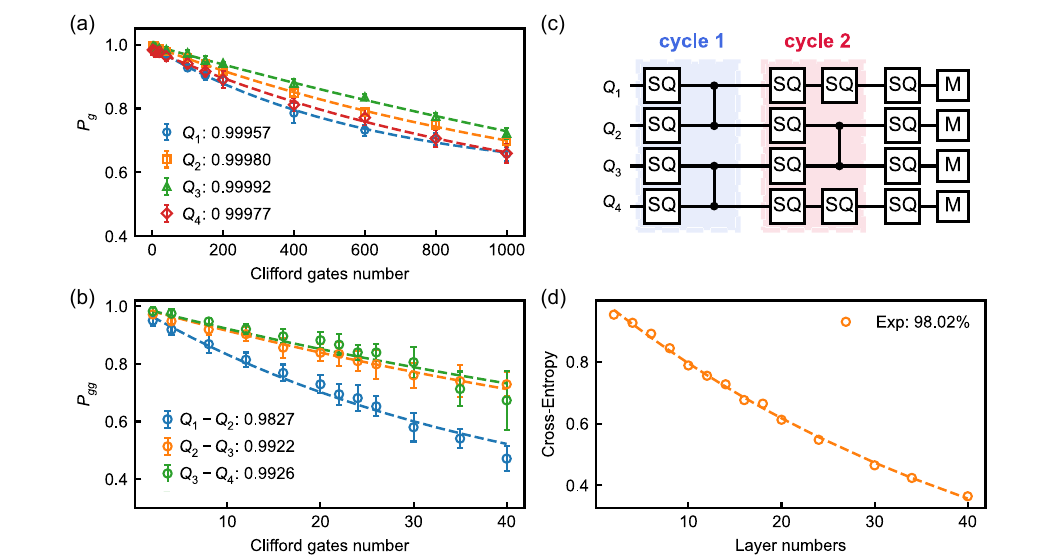} \caption{\label{modular_results}\textbf{Characterization of individual gate and parallel cycle fidelities.} 
    (a) Simultaneous single-qubit RB performed on $Q_1$--$Q_4$. All average single-qubit gate fidelities exceed $99.95\%$. 
    (b) Two-qubit RB decays for the three active CZ pairs, yielding CZ gate fidelities between $98.27\%$ and $99.26\%$. 
    (c) Schematic of the RCS sequence used to assess parallel operation performance, featuring two alternating cycles of single- and two-qubit gates followed by simultaneous measurement. 
    (d) Decay of the cross-entropy versus circuit depth (layer numbers) from the RCS experiment. The exponential fit indicates an average parallel cycle fidelity of $98.02\%$.}
\end{figure*}

\begin{table}[htbp] 
\centering 
\caption{Parameters of modular chip} 
\label{tab:qubit_params} 
\begin{tabular}{@{}lcccc@{}} 
\toprule 
Qubit & Q1 & Q2 & Q3 & Q4 \\ 
\midrule 
$f_{ge}$ (\si{\mega\hertz}) & 277 & 316 & 241 & 267 \\ 
$f_{gf}$ (\si{\giga\hertz}) & 4.91 & 4.92 & 5.77 & 5.00 \\ 
$f_{r}$ (\si{\giga\hertz}) & 5.07 & 5.02 & 5.43 & 5.53 \\ 
$T_1$ (\si{\micro\second}) & 43 & 82 & 129 & 52 \\ 
$T_2^{\text{echo}}$ (\si{\micro\second}) & 15 & 96 & 34 & 46 \\ 
$T_g$ of SQ (\si{\nano\second}) & 20 & 20 & 30 & 30 \\ 
Individual-SQ RB (\%) & 99.973 & 99.990 & 99.992 & 99.974 \\ 
\midrule 

Component & 
\multicolumn{4}{c}{
    \begin{tabular}{@{}ccc@{}}
        $Q_1C_{12}Q_2$ & $Q_2C_{23}Q_3$ & $Q_3C_{34}Q_4$
    \end{tabular}
} \\ 
\midrule 
Coupler-ON bias ($\Phi_0$) & 
\multicolumn{4}{c}{
    \begin{tabular}{@{}ccc@{}}
        \makebox[1.4cm][c]{0.5} & \makebox[1.4cm][c]{0.5} & \makebox[1.4cm][c]{0.5}
    \end{tabular}
} \\ 
Drive state & 
\multicolumn{4}{c}{
    \begin{tabular}{@{}ccc@{}}
        \makebox[1.4cm][c]{$|ggg\rangle$} & \makebox[1.4cm][c]{$|egg\rangle$} & \makebox[1.4cm][c]{$|gge\rangle$}
    \end{tabular}
} \\ 
Microwave duration (\si{\nano\second}) & 
\multicolumn{4}{c}{
    \begin{tabular}{@{}ccc@{}}
        \makebox[1.4cm][c]{60} & \makebox[1.4cm][c]{60} & \makebox[1.4cm][c]{60}
    \end{tabular}
} \\ 
Drive freq (\si{\giga\hertz}) & 
\multicolumn{4}{c}{
    \begin{tabular}{@{}ccc@{}}
        \makebox[1.4cm][c]{4.764} & \makebox[1.4cm][c]{5.157} & \makebox[1.4cm][c]{5.198}
    \end{tabular}
} \\ 
Tow-qubit RB (\%) & 
\multicolumn{4}{c}{
    \begin{tabular}{@{}ccc@{}}
        \makebox[1.4cm][c]{98.27} & \makebox[1.4cm][c]{99.22} & \makebox[1.4cm][c]{99.26}
    \end{tabular}
} \\ 
\bottomrule 
\end{tabular} 
\end{table}

This section provides additional characterization of a four-qubit modular device, which serves as the basic building block of the scalable architecture discussed in the main text (Fig.~2). All tunable couplers are biased at their coupling-OFF positions during idle operation and are activated only during CZ gates via square flux pulses. The parameters for single-qubit operations are summarized in Table~\ref{tab:qubit_params}. Fig.~\ref{modular_results}(a) shows randomized benchmarking (RB) results for simultaneous single-qubit operations on all four qubits. The extracted fidelities are $99.957\%$, $99.980\%$, $99.992\%$, and $99.977\%$ for $Q_1$--$Q_4$, respectively, indicating that the unit-cell design enables high-fidelity parallel control with negligible crosstalk.

For two-qubit gates, the relevant parameters are listed in Table~\ref{tab:qubit_params}. During idle periods, all couplers remain in the coupling-OFF configuration. Entangling operations are realized by applying square flux pulses to selectively activate the target coupler, enabling the microwave-driven transitions described in the main text. Fig.~\ref{modular_results}(b) shows RB results for the three CZ pairs under this on-demand activation scheme, yielding fidelities of $98.27\%$ ($Q_1$--$Q_2$), $99.22\%$ ($Q_2$--$Q_3$), and $99.26\%$ ($Q_3$--$Q_4$).
To further benchmark parallel performance, we perform random circuit sampling (RCS) involving all four qubits, consistent with the protocol in the main text. The sequence, shown in Fig.~\ref{modular_results}(c) , consists of alternating layers of random single-qubit gates and CZ gates, including single-pair configurations ($Q_2$--$Q_3$) and parallel ($Q_1$--$Q_2$ and $Q_3$--$Q_4$) enabled by independent coupler activation.

For an $N$-qubit system ($N=4$ here), each circuit produces a measured probability distribution $\{P_x\}$ over computational basis states $x \in \{g,e\}^N$, which is compared to the ideal distribution $\{P_x^{\mathrm{th}}\}$ obtained from classical simulation. Sequence fidelity (cross entropy) is defined as\cite{boixo2018characterizing}
\begin{equation}
F_{\mathrm{seq}} =
\frac{(M-U)(E-U)}{(E-U)^2},
\end{equation}
where $E = \sum_x \left(P_x^{\mathrm{th}}\right)^2$, $U = \frac{1}{2^N}\sum_x P_x^{\mathrm{th}}$, and $M = \sum_x P_x P_x^{\mathrm{th}}$.
Both the sequence fidelity and purity decay exponentially with circuit depth $L$ as $F_{\mathrm{seq}}(L) = A p^{L} + B$, where $p$ characterizes the effective fidelity per cycle. The corresponding error per cycle (EPC) is given by
\begin{equation}
\mathrm{EPC} = \frac{2^N-1}{2^N}\,(1-p).
\end{equation}
As shown in Fig.~\ref{modular_results}(d), we extract an average cycle fidelity of $98.02\%$ for parallel random circuits in the modular device via exponential fitting. Together with the high-fidelity simultaneous single-qubit operations, these results confirm that the modular unit supports low-crosstalk parallel control and scalable multi-qubit operation, consistent with the conclusions of the main text.


\section{22-Qubit Processor Summary}
\label{22Q}

\begin{figure*}[bt]
  \includegraphics[width = \textwidth]{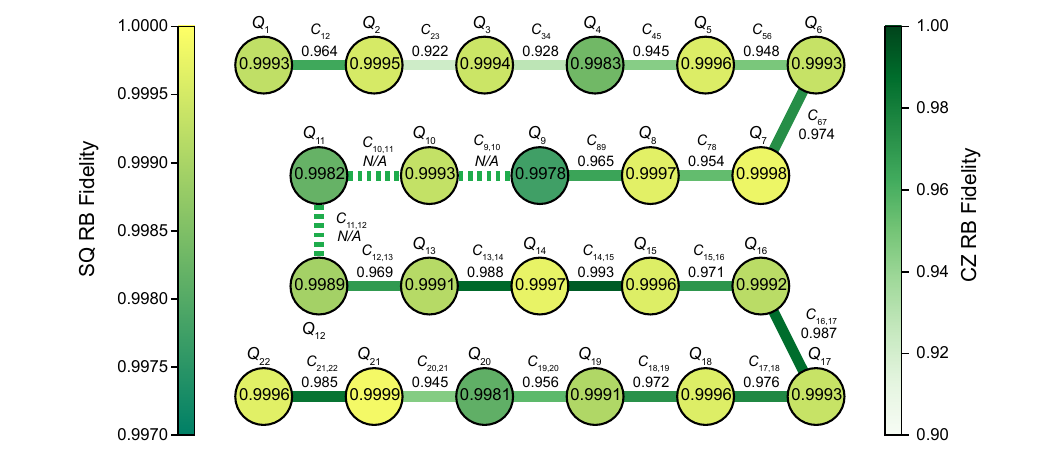} 
  \caption{\textbf{Parallel gate performance map of the 22-qubit processor.} 
    Spatial layout displaying the fidelities of simultaneous single-qubit (SQ) gates and two-qubit CZ gates. 
    The fill colors of the nodes ($Q_1$--$Q_{22}$) and the solid edges correspond to the SQ RB and CZ RB fidelities, respectively, mapped to the left and right colorbars. Explicit fidelity values are annotated adjacent to each component. 
    Inactive coupling paths ($C_{9,10}, C_{10,11}, C_{11,12}$) are indicated by dashed lines.} \label{22Q gate fidelity}
\end{figure*}

\begin{table}[h]
\centering
\caption{Parameters for 22-qubit processor.}
\label{tab:22Q qubit_params}
\setlength{\tabcolsep}{4pt}
\resizebox{0.48\textwidth}{!}{%
\begin{tabular}{c|cccccc}
\toprule
Qubit & $f_r$(GHz) & $f_{ge}$(MHz) & $f_{gf}$(GHz)& $T_1$($\mu$s) & $T_2^R$($\mu$s) & $T_2^{\mathrm{echo}}$($\mu$s) \\
\midrule
$Q_1$  & 5.09 & 297 & 4.79 & 103 & 18 & 26 \\
$Q_2$  & 5.24 & 387 & 4.92 & 111 & 22 & 25 \\
$Q_3$  & 5.49 & 264 & 4.35 & 87 & 12 & 17 \\
$Q_4$  & 5.55 & 122 & 4.61 & 105 & 13 & 14 \\
$Q_5$  & 5.14 & 306 & 5.03 & 92 & 56 & 100 \\
$Q_6$  & 5.19 & 402 & 4.44 & 123 & 21 & 24 \\
$Q_7$  & 5.48 & 282 & 4.56 & 113 & 39 & 120 \\
$Q_8$  & 5.54 & 166 & 4.02 & 38 & 18 & 23 \\
$Q_9$  & 5.08 & 324 & 3.96 & 142 & 34 & 63 \\
$Q_{10}$ & 5.24 & 399 & 4.67 & 113 & 53 & 77 \\
$Q_{11}$ & 5.39 & 293 & 4.15 & 119 & 39 & 98 \\
$Q_{12}$ & 5.54 & 140 & 3.62 & 9 & 8 & 17 \\
$Q_{13}$ & 5.09 & 307 & 3.64 & 87 & 13 & 50 \\
$Q_{14}$ & 5.24 & 337 & 3.96 & 141 & 42 & 125 \\
$Q_{15}$ & 5.49 & 262 & 3.88 & 74 & 6 & 25 \\
$Q_{16}$ & 5.39 & 152 & 4.47 & 18 & 6 & 17 \\
$Q_{17}$ & 5.08 & 293 & 4.47 & 75 & 8 & 14 \\
$Q_{18}$ & 5.23 & 411 & 3.79 & 83 & 16 & 19 \\
$Q_{19}$ & 5.48 & 267 & 3.58 & 126 & 6 & 8 \\
$Q_{20}$ & 5.53 & 123 & 4.30 & 13 & 2 & 3 \\
$Q_{21}$ & 5.13 & 280 & 4.66 & 158 & 5 & 8 \\
$Q_{22}$ & 5.19 & 373 & 5.09 & 203 & 22 & 23 \\
\bottomrule
\end{tabular}%
}
\end{table}
The 22-qubit processor extends the four-qubit modular unit into a linear chain with tunable couplers connecting nearest neighbors. Device parameters, including qubit and resonator frequencies, coherence times, and gate fidelities, are summarized in Fig.~\ref{22Q gate fidelity} and Table~\ref{tab:22Q qubit_params}.
The qubit and resonator frequencies are distributed over a wide range, with readout resonators $f_r$ spanning 5.0--5.6~GHz and qubit transition frequencies $f_{ge}$ ranging from 100 to 400~MHz. This spectral allocation avoids crowding and ensures sufficient frequency selectivity for parallel control, consistent with the design principles described in the main text.

The coherence properties remain robust across the processor, with energy relaxation times $T_1$ reaching up to 203~$\mu$s and averaging above 100~$\mu$s. The dephasing times, including Ramsey ($T_2^R$) and spin-echo ($T_2^{\mathrm{echo}}$), exhibit noticeable variation across the device, indicating non-uniform noise environments. This spread is primarily attributed to fabrication- and materials-related imperfections rather than limitations of the circuit architecture. Further improvements in fabrication processes and materials optimization are expected to enhance coherence uniformity and overall device performance. Parallel single-qubit RB yields average fidelities above $99.9\%$, with the best qubits reaching $99.99\%$, confirming effective suppression of residual couplings during idle operation. For two-qubit gates, the microwave-activated CZ operations are systematically calibrated across the device, with an average fidelity of $95.65\%$ and a maximum of $99.26\%$. 

Overall, single-qubit operations exhibit consistently high performance across the processor, with fidelities exceeding $99.9\%$, demonstrating the effectiveness of the architecture in suppressing residual couplings and enabling high-fidelity parallel single-qubit control. In contrast, the performance of two-qubit gates shows a larger spread, with fidelities currently limited by coherence and dephasing variations across the device. As demonstrated in the main text, optimized CZ gates can reach fidelities above $99.9\%$, indicating that the present limitations are not intrinsic to the architecture but arise from device-level imperfections. These results suggest substantial room for improvement through enhanced coherence and calibration. With continued advances in fabrication and control optimization, the architecture is expected to support uniformly high-fidelity two-qubit operations and improved large-scale entanglement performance.

\section{Characterization of Residual ZZ and XX Couplings}
\label{Couplings}
\begin{figure*}
  \includegraphics[width = 1\textwidth]{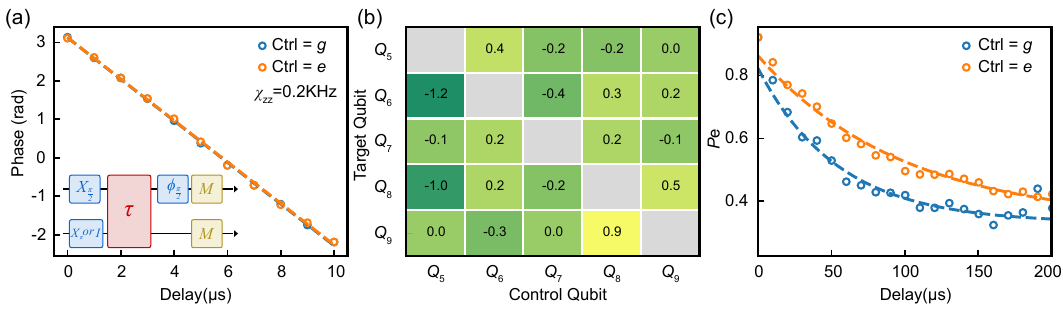} 
  \caption{\textbf{Characterization of residual couplings.} 
    (a), (b) Residual ZZ coupling measured via conditional Ramsey oscillations. 
    (c) XX-induced relaxation characterized by conditional $T_1$ measurements.}
    \label{Condition T1}  
\end{figure*}

We first characterize the ZZ coupling between two qubits via a conditional frequency measurement. By definition, the ZZ interaction corresponds to a state-dependent frequency shift of one qubit conditioned on the state of the other, and can be described by the Hamiltonian $H_{zz} = g_{zz}\,\sigma_z^c \sigma_z^t$. Experimentally, we prepare the control qubit in either $|g\rangle$ or $|e\rangle$ and measure the transition frequency of the target qubit using a standard Ramsey-type experiment. The ZZ coupling is then extracted from the frequency difference
\begin{equation}
g_{zz} = \omega_{t|g} - \omega_{t|e},
\end{equation}
where $\omega_{t|g}$ ($\omega_{t|e}$) denotes the target qubit frequency conditioned on the control qubit being in $|g\rangle$ ($|e\rangle$). In Fig.~\ref{Condition T1}(a), we show the measured Ramsey phase as a function of delay time. The difference in the phase accumulation rates for the control qubit prepared in $|g\rangle$ and $|e\rangle$ directly yields the ZZ coupling strength. 
In this case, the measured ZZ coupling is around $0.2~\mathrm{kHz}$. To provide a direct comparison with the main text, we characterized the ZZ coupling for the same set of qubits while the couplers were biased to the coupling-ON configuration. The results, depicted in Fig.~\ref{Condition T1}(b), show a maximum coupling strength of approximately 1 kHz. Consistent with the main text, this indicates that our system naturally suppresses ZZ coupling to remarkably low levels, a feature to which the unique frequency-separation architecture contributes.

In addition to the longitudinal ZZ coupling, a residual transverse XX coupling may also be present. This coupling can be characterized using a conditional-$T_1$ measurement protocol, which is particularly effective in the weak-coupling regime where coherent excitation exchange between the qubits is overdamped and therefore not directly observable as time-domain oscillations. We consider two resonant qubits coupled via a transverse interaction of the form
\begin{equation}
H = g_{xx}(\sigma_+^c\sigma_-^t + \sigma_-^c\sigma_+^t),
\end{equation}
where $g_{xx}$ denotes the effective XX coupling strength. 
In the presence of dissipation, the system dynamics is governed by a Lindblad master equation with relaxation rates $\kappa_c$ and $\kappa_t$ for the control and target qubits, respectively. 

In the weak-coupling limit $g_{xx} \ll \kappa_c + \kappa_t$, the coherent exchange between the states $|ge\rangle$ and $|eg\rangle$ becomes overdamped. 
In this regime, the fast-decaying coherence between the two states can be adiabatically eliminated, resulting in an effective description where the target qubit acquires an additional decay channel mediated by the control qubit. 
This procedure is formally equivalent to the standard treatment of Purcell decay in cavity QED systems, where a lossy mode induces an additional relaxation channel for a coupled quantum system (see, e.g., Refs.~\cite{purcell1946resonance,raimond2006monitoring,blais2004cavity}). 
The resulting effective relaxation rate of the target qubit, conditioned on the control qubit being in the ground state, is given by
\begin{equation}
\Gamma_{t|g} = \kappa_t + \frac{4g_{xx}^2}{\kappa_c + \kappa_t}.
\end{equation}

By contrast, when the control qubit is initialized in the excited state $|e\rangle$, the transverse interaction is effectively blocked within the two-level approximation, and the target qubit relaxes at its intrinsic rate $\Gamma_{t|e} \approx \kappa_t$. 
Therefore, the difference in the measured decay rates, $\Delta \Gamma = \Gamma_{t|g} - \Gamma_{t|e}$, provides a direct estimate of the residual coupling strength,
\begin{equation}
g_{xx} = \frac{1}{2}\sqrt{(\kappa_c + \kappa_t)\Delta \Gamma}.
\end{equation}

Experimentally, we implement this protocol on the qubit pair $Q_6$--$Q_7$, with $Q_7$ serving as the target qubit. Since the transition frequency of $Q_7$ is lower than that of $Q_6$, $Q_7$ is tuned into resonance with $Q_6$ via flux biasing during the measurement. We then prepare $Q_6$ in either $|g\rangle$ or $|e\rangle$ and measure the corresponding $T_1$ decay of $Q_7$. We observe a clear state dependence: the relaxation of $Q_7$ is faster when $Q_6$ is initialized in $|g\rangle$, and slower when $Q_6$ is initialized in $|e\rangle$. This behavior is consistent with the presence of a finite residual XX coupling that opens an additional decay channel only when the control qubit is in the ground state. As shown in Fig.~\ref{Condition T1}(c), fitting the measured conditional-$T_1$ traces yields a residual XX coupling strength of $g_{xx} \approx 7~\mathrm{kHz}$ between $Q_6$ and $Q_7$.

\section{Estimation of Spectator-Induced Coupling Errors}
\label{Spectator Errors}
To quantify spectator-induced errors during two-qubit gate execution, we examine the behavior of a target qubit pair ($Q_A$, $Q_B$) when neighboring spectator qubits are prepared in various computational states. As discussed in the main text, when the intermediate coupler connecting the target pair to the spectators is biased to the coupling-OFF configuration, both the transition frequency shifts and the resulting controlled-phase errors remain negligible. However, when this intermediate coupler is activated (coupling-ON configuration), the transition frequency of the target pair exhibits a strong dependence on the spectator state. This spectator-induced frequency shift acts as a significant effective detuning, preventing a single parameter set from implementing a consistent CZ gate and precluding a direct experimental extraction of the phase error. 

To quantitatively evaluate this crosstalk, we map the experimentally measured transition frequency shifts to equivalent controlled-phase errors. In the presence of a detuning $\Delta = \omega_d - \omega_d^\prime$ between the calibrated drive frequency $\omega_d$ and the spectator-shifted transition frequency $\omega_d^\prime$, the driven dynamics deviate from an ideal resonant rotation. Within the rotating-wave approximation, the evolution during the CZ gate is governed by the effective Hamiltonian
\begin{equation}
H_{\mathrm{eff}} = \frac{\Delta}{2}\sigma_z + \frac{\Omega_d}{2}\sigma_x,
\end{equation}
where $\Omega_d$ is the resonant Rabi frequency. Under this Hamiltonian, the system undergoes a rotation about an effective tilted axis $\mathbf{n} = (\Omega_d, 0, \Delta)/\Omega_{tot}$, with the generalized Rabi frequency given by $\Omega_{tot} = \sqrt{\Delta^2 + \Omega_d^2}.$
This effective rotation axis forms an angle $\theta$ with the $z$-axis, satisfying $\cos\theta = \Delta / \Omega_{tot}$.

For a complete cyclic trajectory—corresponding to one generalized Rabi cycle of duration $T = 2\pi/\Omega_{tot}$—the state vector traces a closed path on the Bloch sphere, subtending a solid angle of $2\pi(1 - \cos\theta)$. The accumulated geometric phase $\varphi_g$ corresponds to half of this solid angle $\varphi_g=\pi(1 - \cos\theta)$.
The deviation of this accumulated phase from the ideal $\pi$ directly reflects the phase error induced by the detuning. In our analysis, the spectator-induced frequency shift directly dictates the effective detuning $\Delta$ during the gate. By incorporating these measured shifts into numerical simulations of the cyclic trajectory, we extract the deviation of the accumulated controlled phase from $\pi$. This methodology enables us to translate observed spectral shifts into equivalent phase errors, providing a direct and quantitative measure of spectator-induced crosstalk when couplers are activated across the processor architecture.

\section{CZ Calibration}
\label{CZ}
\begin{figure}
  \includegraphics[width = 0.48\textwidth]{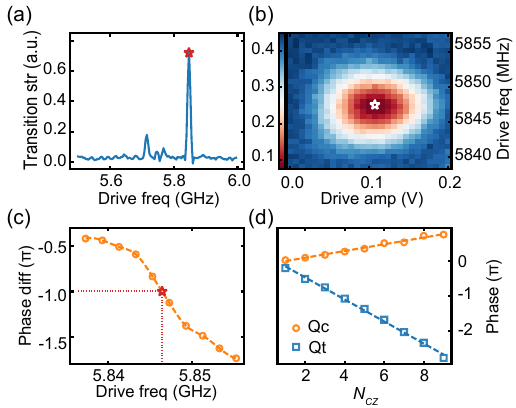} \caption{\textbf{Calibration sequence for the microwave-activated CZ gate.} 
    (a) Coupler spectroscopy performed with qubits in the $\ket{ee}$ state to locate an isolated transition suitable for the CZ operation. 
    (b) 2D Rabi chevron measurement scanning the coupler drive frequency and amplitude. This identifies the resonance condition and parameters required for a complete population return. 
    (c) Conditional phase measurement via Ramsey interferometry. The drive frequency is fine-tuned to yield exactly a $\pi$ phase difference between the target qubit trajectories when the control qubit is in $\ket{g}$ versus $\ket{e}$. 
    (d) Measurement of the accumulated dynamical phase on individual qubits under repeated CZ gates. The extracted per-gate phase offsets from linear fits are compensated using virtual-$Z$ gates.}\label{FigS5} 
\end{figure}

We calibrate each CZ gate in the FTF chain using a four-step protocol: (i) identification of the relevant coupler-assisted transition, (ii) determination of the optimal drive amplitude and frequency for coherent population exchange, (iii) calibration of the conditional $\pi$ phase, and (iv) compensation of residual single-qubit phase shifts. Representative data are shown in Fig.~\ref{FigS5}.

We first identify candidate transitions by preparing the qubits in the $\ket{ee}$ state (as a representative initial condition) and applying a frequency-swept microwave drive to the FTF system. As shown in Fig.~\ref{FigS5}(a), multiple resonances corresponding to transitions within the hybridized manifold are observed. We select an isolated, high-contrast feature as the target transition to minimize leakage and reduce sensitivity to spectral crowding. Next, we perform a two-dimensional Rabi chevron measurement around the selected transition by sweeping the drive frequency and amplitude at a fixed pulse duration of 160~ns. As shown in Fig.~\ref{FigS5}(b), the measured population exhibits coherent Rabi oscillations, forming a characteristic chevron pattern. From the chevron center and fitted oscillation envelope, we extract the optimal drive frequency and amplitude corresponding to a full population exchange.

To implement a CZ gate, we convert this population-exchange condition into a conditional phase operation. The target qubit is prepared in a superposition $(\ket{g}+\ket{e})$, while the control qubit is initialized in $\ket{g}$ and $\ket{e}$ in separate runs. Sweeping the drive frequency near the resonance, we measure the accumulated relative phase. As shown in Fig.~\ref{FigS5}(c), the CZ condition is identified when the conditional phase reaches $\pi$. Finally, we quantify and compensate residual single-qubit phase shifts arising from ac-Stark effects. This is achieved by applying repeated CZ gates and extracting the accumulated phase $\varphi(N_{CZ})$ on the target qubit. A linear fit $\varphi(N_{CZ})=N_{CZ}\cdot\varphi_{sq}+\varphi_0$ yields the per-gate phase offset $\varphi_{sq}$, which is compensated using virtual-$Z$ rotations in subsequent control pulses, as shown in Fig.~\ref{FigS5}(d). This calibration protocol yields a robust and reproducible CZ gate and is directly compatible with parallel implementation across multiple couplers in the scalable FTF architecture.

\section{Experimental Generation of GHZ States and Error Mitigation}
\label{GHZ}
In this section, we detail the experimental sequence used to generate GHZ states, the methodology for extracting state fidelity via parity oscillations, and the error mitigation protocols employed to suppress initialization and readout errors.

\subsection{GHZ State Preparation}

Following the sequence illustrated in Fig.~\ref{FigS4}, we prepare an $N$-qubit GHZ state $\ket{\mathrm{GHZ}}=(\ket{ggg\dots g}+\ket{eee\dots e})/\sqrt{2}$. As shown in the sequence, the system is first initialized to the ground state by a flux pulse (labeled $\text{R}$), using a protocol similar to Ref.~\cite{bao2022fluxonium}. Starting from this state, a Hadamard-type rotation prepares the first qubit in an equal superposition. Successive CNOT operations then distribute this coherence across the FTF chain. In the experiment, each CNOT gate is implemented using a calibrated CZ gate combined with three single-qubit rotations. To maximize parallelism in the linear chain, entanglement is generated starting from the central qubit and propagating outward to both sides simultaneously, yielding a total CZ circuit depth of $\lceil N/2 \rceil$. Ideally, the generated density matrix can be written as
\begin{equation}
\begin{split}
\rho_{\mathrm{GHZ}} =\;& \mathrm{Re}(\rho_{gg\dots g,ee\dots e}) \left( |gg\dots g\rangle\langle ee\dots e| + \mathrm{h.c.} \right) \\
& + 
\frac{1}{2} \left( |g\rangle\langle g|^{\otimes N} + |e\rangle\langle e|^{\otimes N} \right),
\end{split}
\end{equation}
where the off-diagonal element $\rho_{gg\dots g,ee\dots e}$ characterizes the global multi-qubit coherence. To extract this coherence experimentally, we perform parity oscillations. We apply a collective phase rotation $R_Z^N=\bigotimes_{i=1}^{N} R_z^{(i)}(\phi)$, implemented as virtual-$Z$ frame updates. This imprints a relative phase difference between the two components:
\begin{equation}
\begin{aligned}
|gg\dots g\rangle &\rightarrow e^{-iN\phi/2}|gg\dots g\rangle, \\
|ee\dots e\rangle &\rightarrow e^{+iN\phi/2}|ee\dots e\rangle,
\end{aligned}
\end{equation}
resulting in a relative phase accumulation of $N\phi$. Following a global $X_{\pi/2}$ rotation, the parity operator $\Pi_N = \bigotimes_{i=1}^{N} Z_i$ is measured. The expected parity signal $P(\phi) = \langle \Pi_N(\phi) \rangle = \sum_x(-1)^{\omega(x)}P_x(\phi)$, where $\omega(x)$ is the Hamming weight of bitstring $x$, directly probes the coherence
\begin{equation}
P(\phi) = 2\,\mathrm{Re}(\rho_{gg\dots g,ee\dots e}) \cos(N\phi+\phi_0).
\label{eq:ideal_parity}
\end{equation}
Thus, the ideal parity oscillation exhibits a single harmonic at order $N$. Fitting the experimental data yields the oscillation amplitude $A_N$, from which the coherence term is estimated as $\mathrm{Re}(\rho_{gg\dots g,ee\dots e}) \approx A_N/2$ \cite{song201710, omran2019generation, bao2024creating}. The total GHZ fidelity is then obtained using the corrected $Z$-basis populations alongside the extracted coherence amplitude:
\begin{equation}
\mathcal{F} = \frac{\rho_{gg\dots g, gg\dots g}+\rho_{ee\dots e,ee\dots e}}{2}+\mathrm{Re}(\rho_{gg\dots g,ee\dots e}).
\end{equation}

\subsection{Effect of Initialization Errors and Pre-selection Protocol}

Initialization errors become increasingly significant for larger GHZ states because a single residual excitation can corrupt the global parity signal. To understand this, consider the effect of imperfect initialization on one qubit, labeled $Q_k$. The initial state is modeled as a classical mixture $\rho_{\mathrm{init}}^{(k)} = (1-\epsilon)|g\rangle\langle g| + \epsilon |e\rangle\langle e|$. After the GHZ circuit, the system evolves into a mixture of two orthogonal GHZ-like branches:
\begin{equation}
\rho = (1-\epsilon)\rho_{\mathrm{GHZ}}^{(0)} + \epsilon\,\rho_{\mathrm{GHZ}}^{(1)},
\end{equation}
where $\rho_{\mathrm{GHZ}}^{(0)}$ corresponds to the ideal GHZ state, and $\rho_{\mathrm{GHZ}}^{(1)}$ corresponds to a state where $Q_k$ is erroneously flipped. 

These two branches accumulate different phases and contribute distinct oscillation components to the parity signal. As a result, the measured parity takes the general multi-frequency form
\begin{equation}
P(\phi) = A_N \cos(N\phi) + \sum_{m<N} A_m \cos(m\phi),
\label{eq:multi_freq}
\end{equation}
where the additional lower-order harmonics ($m<N$) arise directly from incoherent contributions associated with incorrect initialization. Physically, initialization errors reduce the weight of the coherent GHZ component while introducing spurious oscillation frequencies, leading to reduced contrast and multi-frequency beating in the parity signal.

\begin{figure}
  \includegraphics[width = 0.48\textwidth]{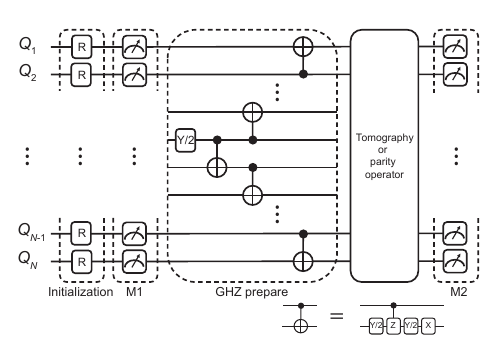}
  \caption{\textbf{GHZ state preparation.} Circuit diagram for generating an $N$-qubit GHZ state on a 1D chain. To suppress initialization errors, a pre-selection measurement (M1) is performed immediately after the Reset (R) gate. The entanglement is generated starting from the central qubit and distributed symmetrically via a sequence of CNOT gates. Parallel operations are utilized to reduce the overall circuit depth. The bottom inset illustrates the compilation of a CNOT gate using the native CZ gate and single-qubit rotations. Prior to the final measurement (M2), operations for parity oscillation or state tomography are applied to characterize the GHZ coherence.} 
  \label{FigS4}  
\end{figure}

To suppress these initialization-induced errors, we employ a quantum non-demolition (QND) pre-selection protocol \cite{heinsoo2018rapid}. As depicted in Fig.~\ref{FigS4}, an initial measurement (M1) is performed immediately after state initialization, followed by the GHZ circuit and a final measurement (M2). Experimental shots are retained \textit{only} when the M1 outcome corresponds to the all-ground-state configuration. This procedure projects the system onto the subspace consistent with correct initialization, effectively filtering out the erroneous $\rho_{\mathrm{GHZ}}^{(1)}$ branches. Consequently, the parity signal recovers the single-harmonic form described in Eq.~(\ref{eq:ideal_parity}) without observable lower-order frequency components. While imperfect readout fidelity allows a small fraction of excited states to pass the filter (consistent with numerical simulations), the pre-selection procedure substantially stabilizes the observed oscillation contrast.

\subsection{Readout Error Mitigation and Fidelity Analysis}

In addition to pre-selection, measurement errors in M2 are further reduced through multi-qubit readout calibration based on a confusion-matrix correction. The full assignment matrix is constructed from measurements of computational basis states. Corrected probability distributions are obtained using a constrained maximum-likelihood estimator that enforces positivity and normalization \cite{aasen2024readout}. All GHZ data reported in this work include both pre-selection filtering and readout correction, ensuring that extracted observables primarily reflect gate performance rather than measurement or initialization bias.

Ultimately, the measured GHZ fidelities agree well with predictions based on independently characterized gate fidelities $F_{\text{theory}} =  F_{\text{SQ}} \left(Q_j\right)\cdot \prod_{i=1}^{N-1}\left(F_{\text{SQ}}\left(Q_i\right)\right)^{3} \cdot\prod_{i=1}^{N-1}F_{\text{CZ}}\left(C_{i,{i+1}}\right)$ (Fig.~4(b) in main text), indicating that coherent gate errors are the primary limitation. Decoherence contributes only weakly; for instance, the total duration of the 10-qubit GHZ circuit is 1.064~$\mu$s, which is much shorter than the average Ramsey dephasing time of 15~$\mu$s. We note that the measured GHZ fidelities slightly exceed the product of average single- and two-qubit RB gate fidelities. This discrepancy arises because RB averages over random Clifford inputs, whereas the GHZ sequence predominantly addresses equatorial states. State-dependent coherent errors can therefore affect RB averages more strongly than the fixed operations in the GHZ circuit, leading to a slight underestimation of the sequence fidelity by RB.



\bibliographystyle{apsrev4-2}
\bibliography{fluxonium_refs}

@article{Kjaergaard2020,
   author = "Kjaergaard, Morten and Schwartz, Mollie E. and Braumüller, Jochen and Krantz, Philip and Wang, Joel I.-J. and Gustavsson, Simon and Oliver, William D.",
   title = "Superconducting Qubits: Current State of Play", 
   journal= "Annu. Rev. Condens. Matter Phys.",
   year = "2020",
   volume = "11",
   number = "Volume 11, 2020",
   pages = "369-395",
   doi = "https://doi.org/10.1146/annurev-conmatphys-031119-050605",
   url = "https://www.annualreviews.org/content/journals/10.1146/annurev-conmatphys-031119-050605",
   publisher = "Annual Reviews",
   issn = "1947-5462",
   type = "Journal Article",
  }

@article{arute2019quantum,
  title={Quantum Supremacy Using a Programmable Superconducting Processor},
  author={Arute, Frank and Arya, Kunal and Babbush, Ryan and Bacon, Dave and Bardin, Joseph C and Barends, Rami and Biswas, Rupak and Boixo, Sergio and Brandao, Fernando GSL and Buell, David A. and others},
  journal={Nature},
  volume={574},
  number={7779},
  pages={505--510},
  year={2019},
  publisher={Nature Publishing Group UK London},
  doi={https://doi.org/10.1038/s41586-019-1666-5}
}

@article{Wu2021,
  title = {Strong Quantum Computational Advantage Using a Superconducting Quantum Processor},
  author = {Wu, Yulin and Bao, Wan-Su and Cao, Sirui and Chen, Fusheng and Chen, Ming-Cheng and Chen, Xiawei and Chung, Tung-Hsun and Deng, Hui and Du, Yajie and Fan, Daojin and Gong, Ming and Guo, Cheng and Guo, Chu and Guo, Shaojun and Han, Lianchen and Hong, Linyin and Huang, He-Liang and Huo, Yong-Heng and Li, Liping and Li, Na and others},
  journal = {Phys. Rev. Lett.},
  volume = {127},
  issue = {18},
  pages = {180501},
  numpages = {7},
  year = {2021},
  month = {Oct},
  publisher = {American Physical Society},
  doi = {10.1103/PhysRevLett.127.180501},
  url = {https://link.aps.org/doi/10.1103/PhysRevLett.127.180501}
}

@Article{Abanin2025,
author={{Google Quantum AI and Collaborators}},
title={Observation of Constructive Interference at the Edge of Quantum Ergodicity},
journal={Nature},
year={2025},
month={Oct},
day={01},
volume={646},
number={8086},
pages={825-830},
issn={1476-4687},
doi={10.1038/s41586-025-09526-6},
url={https://doi.org/10.1038/s41586-025-09526-6}
}

@article{kim2023evidence,
  title={Evidence for the Utility of Quantum Computing Before Fault Tolerance},
  author={Kim, Youngseok and Eddins, Andrew and Anand, Sajant and Wei, Ken Xuan and Van Den Berg, Ewout and Rosenblatt, Sami and Nayfeh, Hasan and Wu, Yantao and Zaletel, Michael and Temme, Kristan and others},
  journal={Nature},
  volume={618},
  number={7965},
  pages={500--505},
  year={2023},
  publisher={Nature Publishing Group UK London},
  doi = {https://doi.org/10.1038/s41586-023-06096-3}
}

@Article{Jin2025,
author={Jin, Feitong
and Jiang, Si
and Zhu, Xuhao
and Bao, Zehang
and Shen, Fanhao
and Wang, Ke
and Zhu, Zitian
and Xu, Shibo
and Song, Zixuan
and Chen, Jiachen
and Tan, Ziqi
and Wu, Yaozu
and Zhang, Chuanyu
and Gao, Yu
and Wang, Ning
and Zou, Yiren
and Zhang, Aosai
and Li, Tingting
and Zhong, Jiarun
and Cui, Zhengyi
and others},
title={Topological Prethermal Strong Zero Modes on Superconducting Processors},
journal={Nature},
year={2025},
month={Sep},
day={01},
volume={645},
number={8081},
pages={626-632},
issn={1476-4687},
doi={10.1038/s41586-025-09476-z},
url={https://doi.org/10.1038/s41586-025-09476-z}
}

@Article{Acharya2025,
author={{Google Quantum AI and Collaborators}},
title={Quantum Error Correction Below the Surface Code Threshold},
journal={Nature},
year={2025},
month={Feb},
day={01},
volume={638},
number={8052},
pages={920-926},
issn={1476-4687},
doi={10.1038/s41586-024-08449-y},
url={https://doi.org/10.1038/s41586-024-08449-y}
}

@Article{Wang2026,
author={Wang, Ke
and Lu, Zhide
and Zhang, Chuanyu
and Liu, Gongyu
and Chen, Jiachen
and Wang, Yanzhe
and Wu, Yaozu
and Xu, Shibo
and Zhu, Xuhao
and Jin, Feitong
and Gao, Yu
and Tan, Ziqi
and Cui, Zhengyi
and Wang, Ning
and Zou, Yiren
and Zhang, Aosai
and Li, Tingting
and Shen, Fanhao
and Zhong, Jiarun
and Bao, Zehang
and others},
title={Demonstration of Low-Overhead Quantum Error Correction Codes},
journal={Nat. Phys.},
year={2026},
month={Feb},
day={01},
volume={22},
number={2},
pages={308-314},
issn={1745-2481},
doi={10.1038/s41567-025-03157-4},
url={https://doi.org/10.1038/s41567-025-03157-4}
}

@Article{Besedin2026,
author={Besedin, Ilya
and Kerschbaum, Michael
and Knoll, Jonathan
and Hesner, Ian
and B{\"o}deker, Lukas
and Colmenarez, Luis
and Hofele, Luca
and Lacroix, Nathan
and Hellings, Christoph
and Swiadek, Fran{\c{c}}ois
and Flasby, Alexander
and Bahrami Panah, Mohsen
and Colao Zanuz, Dante
and M{\"u}ller, Markus
and Wallraff, Andreas},
title={Lattice Surgery Realized on Two Distance-Three Repetition Codes with Superconducting Qubits},
journal={Nat. Phys.},
year={2026},
month={Feb},
day={01},
volume={22},
number={2},
pages={189-194},
issn={1745-2481},
doi={10.1038/s41567-025-03090-6},
url={https://doi.org/10.1038/s41567-025-03090-6}
}

@article{He2025,
  title = {Experimental Quantum Error Correction below the Surface Code Threshold via All-Microwave Leakage Suppression},
  author = {He, Tan and Lin, Weiping and Wang, Rui and Li, Yuan and Bei, Jiahao and Cai, Jianbin and Cao, Sirui and Chen, Danning and Chen, Kefu and Chen, Xiawei and Chen, Zhe and Chen, Zhiyuan and Chen, Zihua and Chu, Wenhao and Deng, Hui and Ding, Xun and Ding, Zhuzhengqi and Fan, Bo and Fan, Daojin and Fu, Yuanhao and others},
  journal = {Phys. Rev. Lett.},
  volume = {135},
  issue = {26},
  pages = {260601},
  numpages = {7},
  year = {2025},
  month = {Dec},
  publisher = {American Physical Society},
  doi = {10.1103/rqkg-dw31},
  url = {https://link.aps.org/doi/10.1103/rqkg-dw31}
}

@article{Koch2007,
  title={Charge-insensitive Qubit Design Derived from the Cooper Pair Box},
  author={Koch, Jens and Yu, Terri M and Gambetta, Jay and Houck, Andrew A and Schuster, David I and Majer, Johannes and Blais, Alexandre and Devoret, Michel H and Girvin, Steven M and Schoelkopf, Robert J},
  journal={Phys. Rev. A},
  volume={76},
  number={4},
  pages={042319},
  year={2007},
  doi = {https://doi.org/10.1103/PhysRevA.76.042319}
}

@article{Somoroff2023,
  title = {Millisecond Coherence in a Superconducting Qubit},
  author = {Somoroff, Aaron and Ficheux, Quentin and Mencia, Raymond A. and Xiong, Haonan and Kuzmin, Roman and Manucharyan, Vladimir E.},
  journal = {Phys. Rev. Lett.},
  volume = {130},
  issue = {26},
  pages = {267001},
  numpages = {6},
  year = {2023},
  month = {Jun},
  publisher = {American Physical Society},
  doi = {10.1103/PhysRevLett.130.267001},
  url = {https://link.aps.org/doi/10.1103/PhysRevLett.130.267001}
}

@article{Wang2025,
  title = {High-Coherence Fluxonium Qubits Manufactured with a Wafer-Scale-Uniformity Process},
  author = {Wang, Fei and Lu, Kannan and Zhan, Huijuan and Ma, Lu and Wu, Feng and Sun, Hantao and Deng, Hao and Bai, Yang and Bao, Feng and Chang, Xu and Gao, Ran and Gao, Xun and Gong, Guicheng and Hu, Lijuan and Hu, Ruizi and Ji, Honghong and Ma, Xizheng and Mao, Liyong and Song, Zhijun and Tang, Chengchun and others},
  journal = {Phys. Rev. Appl.},
  volume = {23},
  issue = {4},
  pages = {044064},
  numpages = {14},
  year = {2025},
  month = {Apr},
  publisher = {American Physical Society},
  doi = {10.1103/PhysRevApplied.23.044064},
  url = {https://link.aps.org/doi/10.1103/PhysRevApplied.23.044064}
}

@article{Ficheux2021,
  title = {Fast Logic with Slow Qubits: Microwave-Activated Controlled-Z Gate on Low-Frequency Fluxoniums},
  author = {Ficheux, Quentin and Nguyen, Long B. and Somoroff, Aaron and Xiong, Haonan and Nesterov, Konstantin N. and Vavilov, Maxim G. and Manucharyan, Vladimir E.},
  journal = {Phys. Rev. X},
  volume = {11},
  issue = {2},
  pages = {021026},
  numpages = {16},
  year = {2021},
  month = {May},
  publisher = {American Physical Society},
  doi = {10.1103/PhysRevX.11.021026},
  url = {https://link.aps.org/doi/10.1103/PhysRevX.11.021026}
}

@article{bao2022fluxonium,
    title = {Fluxonium: An Alternative Qubit Platform for High-Fidelity Operations},
  author = {Bao, Feng and Deng, Hao and Ding, Dawei and Gao, Ran and Gao, Xun and Huang, Cupjin and Jiang, Xun and Ku, Hsiang-Sheng and Li, Zhisheng and Ma, Xizheng and Ni, Xiaotong and Qin, Jin and Song, Zhijun and Sun, Hantao and Tang, Chengchun and Wang, Tenghui and Wu, Feng and Xia, Tian and Yu, Wenlong and Zhang, Fang and others},
  journal = {Phys. Rev. Lett.},
  volume = {129},
  issue = {1},
  pages = {010502},
  numpages = {6},
  year = {2022},
  month = {Jun},
  publisher = {American Physical Society},
  doi = {10.1103/PhysRevLett.129.010502},
  url = {https://link.aps.org/doi/10.1103/PhysRevLett.129.010502}
}

@article{ma2023native,
  title = {Native Approach to Controlled-{$Z$} Gates in Inductively Coupled Fluxonium Qubits},
  author = {Ma, Xizheng and Zhang, Gengyan and Wu, Feng and Bao, Feng and Chang, Xu and Chen, Jianjun and Deng, Hao and Gao, Ran and Gao, Xun and Hu, Lijuan and Ji, Honghong and Ku, Hsiang-Sheng and Lu, Kannan and Ma, Lu and Mao, Liyong and Song, Zhijun and Sun, Hantao and Tang, Chengchun and Wang, Fei and Wang, Hongcheng and others},
  journal = {Phys. Rev. Lett.},
  volume = {132},
  issue = {6},
  pages = {060602},
  numpages = {6},
  year = {2024},
  month = {Feb},
  publisher = {American Physical Society},
  doi = {10.1103/PhysRevLett.132.060602},
  url = {https://link.aps.org/doi/10.1103/PhysRevLett.132.060602}
}

@article{Lin2025,
  title = {24 Days-Stable CNOT Gate on Fluxonium Qubit with over 99.9\% Fidelity
          },
  author = {Lin, Wei-Ju and Cho, Hyunheung and Chen, Yinqi and Vavilov, Maxim G. and Wang, Chen and Manucharyan, Vladimir E.},
  journal = {PRX Quantum},
  volume = {6},
  issue = {1},
  pages = {010349},
  numpages = {20},
  year = {2025},
  month = {Mar},
  publisher = {American Physical Society},
  doi = {10.1103/PRXQuantum.6.010349}
}

@article{ding2023high,
  title = {High-Fidelity, Frequency-Flexible Two-Qubit Fluxonium Gates with a Transmon Coupler},
  author = {Ding, Leon and Hays, Max and Sung, Youngkyu and Kannan, Bharath and An, Junyoung and Di Paolo, Agustin and Karamlou, Amir H. and Hazard, Thomas M. and Azar, Kate and Kim, David K. and Niedzielski, Bethany M. and Melville, Alexander and Schwartz, Mollie E. and Yoder, Jonilyn L. and Orlando, Terry P. and Gustavsson, Simon and Grover, Jeffrey A. and Serniak, Kyle and Oliver, William D.},
  journal = {Phys. Rev. X},
  volume = {13},
  issue = {3},
  pages = {031035},
  numpages = {24},
  year = {2023},
  month = {Sep},
  publisher = {American Physical Society},
  doi = {10.1103/PhysRevX.13.031035},
  url = {https://link.aps.org/doi/10.1103/PhysRevX.13.031035}
}

@article{moskalenko2022high,
  title={High Fidelity Two-Qubit Gates on Fluxoniums Using a Tunable Coupler},
  author={Moskalenko, Ilya N and Simakov, Ilya A and Abramov, Nikolay N and Grigorev, Alexander A and Moskalev, Dmitry O and Pishchimova, Anastasiya A and Smirnov, Nikita S and Zikiy, Evgeniy V and Rodionov, Ilya A and Besedin, Ilya S},
  journal={npj Quantum Inf.},
  volume={8},
  number={1},
  pages={130},
  year={2022},
  publisher={Nature Publishing Group UK London},
  doi={10.1038/s41534-022-00644-x}
}

@article{zhang2024tunable,
  title={Tunable Inductive Coupler for High-Fidelity Gates Between Fluxonium Qubits},
  author={Zhang, Helin and Ding, Chunyang and Weiss, DK and Huang, Ziwen and Ma, Yuwei and Guinn, Charles and Sussman, Sara and Chitta, Sai Pavan and Chen, Danyang and Houck, Andrew A and others},
  journal={PRX Quantum},
  volume={5},
  number={2},
  pages={020326},
  year={2024},
  publisher={APS},
  doi = {https://doi.org/10.1103/PRXQuantum.5.020326}
}

@unpublished{manucharyan2009coherent,
      title={Coherent Oscillations Between Classically Separable Quantum States of a Superconducting Loop}, 
      author={Manucharyan, Vladimir E and Koch, Jens and Brink, Markus and Glazman, Leonid I and Devoret, Michel H},
      year={2019},
      eprint={0910.3039},
      archivePrefix={arXiv},
      url={https://arxiv.org/pdf/0910.3039}
}

@article{barends2014surface,
  author       = {Barends, Rami and Kelly, Julian and Megrant, Anthony and Veitia, Andrzej and Sank, Daniel and Jeffrey, Evan and White, Ted C and Mutus, Josh and Fowler, Austin G and Campbell, Brooks and others},
  title        = {Superconducting Quantum Circuits at the Surface Code Threshold for Fault Tolerance},
  journal      = {Nature},
  volume       = {508},
  pages        = {500--503},
  year         = {2014},
  doi          = {https://doi.org/10.1038/nature13171}
}

@article{bao2024creating,
  author       = {Bao, Zehang and Xu, Shibo and Song, Zixuan and Wang, Ke and Xiang, Liang and Zhu, Zitian and Chen, Jiachen and Jin, Feitong and Zhu, Xuhao and Gao, Yu and others},
  title        = {Creating and Controlling Global Greenberger-Horne-Zeilinger Entanglement on Quantum Processors},
  journal      = {Nat. Commun.},
  volume       = {15},
  pages        = {8823},
  year         = {2024},
  doi          = {10.1038/s41467-024-53140-5}
}

@article{Reed2010fast,
    author = {Reed, M. D. and Johnson, B. R. and Houck, A. A. and DiCarlo, L. and Chow, J. M. and Schuster, D. I. and Frunzio, L. and Schoelkopf, R. J.},
    title = "{Fast Reset and Suppressing Spontaneous Emission of a Superconducting Qubit}",
    journal = {Appl. Phys. Lett.},
    volume = {96},
    number = {20},
    pages = {203110},
    year = {2010},
    month = {05},
    issn = {0003-6951},
    doi = {10.1063/1.3435463}
}

@article{mcewen2021removing,
  title={Removing Leakage-Induced Correlated Errors in Superconducting Quantum Error Correction},
  author={McEwen, Matt and Kafri, Dvir and Chen, Z and Atalaya, Juan and Satzinger, KJ and Quintana, Chris and Klimov, Paul Victor and Sank, Daniel and Gidney, C and Fowler, AG and others},
  journal={Nat. Commun.},
  volume={12},
  number={1},
  pages={1761},
  year={2021},
  publisher={Nature Publishing Group UK London},
  doi={10.1038/s41467-021-21982-y}
}

@article{Chen2023Compiling,
  title = {Compiling Arbitrary Single-Qubit Gates via the Phase Shifts of Microwave Pulses},
  author = {Chen, Jianxin and Ding, Dawei and Huang, Cupjin and Ye, Qi},
  journal = {Phys. Rev. Res.},
  volume = {5},
  issue = {2},
  pages = {L022031},
  numpages = {6},
  year = {2023},
  month = {May},
  publisher = {American Physical Society},
  doi = {10.1103/PhysRevResearch.5.L022031},
  url = {https://link.aps.org/doi/10.1103/PhysRevResearch.5.L022031}
}

@article{song201710,
  title={10-Qubit Entanglement and Parallel Logic Operations with a Superconducting Circuit},
  author={Song, Chao and Xu, Kai and Liu, Wuxin and Yang, Chui-Ping and Zheng, Shi-Biao and Deng, Hui and Xie, Qiwei and Huang, Keqiang and Guo, Qiujiang and Zhang, Libo and others},
  journal={Phys. Rev. Lett.},
  volume={119},
  number={18},
  pages={180511},
  year={2017},
  publisher={APS},
  doi={https://doi.org/10.1103/PhysRevLett.119.180511}
}

@article{wei2020verifying,
  title={Verifying Multipartite Entangled Greenberger-Horne-Zeilinger States via Multiple Quantum Coherences},
  author={Wei, Ken X and Lauer, Isaac and Srinivasan, Srikanth and Sundaresan, Neereja and McClure, Douglas T and Toyli, David and McKay, David C and Gambetta, Jay M and Sheldon, Sarah},
  journal={Phys. Rev. A},
  volume={101},
  number={3},
  pages={032343},
  year={2020},
  publisher={APS},
  doi = {https://doi.org/10.1103/PhysRevA.101.032343}
}

@article{dicarlo2010preparation,
  title={Preparation and Measurement of Three-Qubit Entanglement in a Superconducting Circuit},
  author={DiCarlo, Leonardo and Reed, Matthew D and Sun, Luyan and Johnson, Blake R and Chow, Jerry M and Gambetta, Jay M and Frunzio, Luigi and Girvin, Steven M and Devoret, Michel H and Schoelkopf, Robert J},
  journal={Nature},
  volume={467},
  number={7315},
  pages={574--578},
  year={2010},
  publisher={Nature Publishing Group UK London},
  doi={https://doi.org/10.1038/nature09416}
}

@article{magesan2012efficient,
  title={Efficient Measurement of Quantum Gate Error by Interleaved Randomized Benchmarking},
  author={Magesan, Easwar and Gambetta, Jay M and Johnson, Blake R and Ryan, Colm A and Chow, Jerry M and Merkel, Seth T and Da Silva, Marcus P and Keefe, George A and Rothwell, Mary B and Ohki, Thomas A and others},
  journal={Phys. Rev. Lett.},
  volume={109},
  number={8},
  pages={080505},
  year={2012},
  publisher={APS},
  doi={https://doi.org/10.1103/PhysRevLett.109.080505}
}

@unpublished{valles2025,
      title={Optimizing the Frequency Positioning of Tunable Couplers in a Circuit QED Processor to Mitigate Spectator Effects on Quantum Operations}, 
      author={S. Vallés-Sanclemente and T. H. F. Vroomans and T. R. van Abswoude and F. Brulleman and T. Stavenga and S. L. M. van der Meer and Y. Xin and A. Lawrence and V. Singh and M. A. Rol and L. DiCarlo},
      year={2025},
      eprint={2503.13225},
      archivePrefix={arXiv},
      url={https://arxiv.org/abs/2503.13225}, 
}

@Article{Klimov2024,
author={Klimov, Paul V.
and Bengtsson, Andreas
and Quintana, Chris
and Bourassa, Alexandre
and Hong, Sabrina
and Dunsworth, Andrew
and Satzinger, Kevin J.
and Livingston, William P.
and Sivak, Volodymyr
and Niu, Murphy Yuezhen
and Andersen, Trond I.
and Zhang, Yaxing
and Chik, Desmond
and Chen, Zijun
and Neill, Charles
and Erickson, Catherine
and Grajales Dau, Alejandro
and Megrant, Anthony
and Roushan, Pedram
and Korotkov, Alexander N. 
and others},
title={Optimizing Quantum Gates Towards the Scale of Logical Qubits},
journal={Nat. Commun.},
year={2024},
month={Mar},
day={18},
volume={15},
number={1},
pages={2442},
issn={2041-1723},
doi={10.1038/s41467-024-46623-y},
url={https://doi.org/10.1038/s41467-024-46623-y}
}

@article{yan2018tunable,
  title={Tunable Coupling Scheme for Implementing High-Fidelity Two-Qubit Gates},
  author={Yan, Fei and Krantz, Philip and Sung, Youngkyu and Kjaergaard, Morten and Campbell, Daniel L and Orlando, Terry P and Gustavsson, Simon and Oliver, William D},
  journal={Phys. Rev. Applied},
  volume={10},
  number={5},
  pages={054062},
  year={2018},
  publisher={APS},
  doi={https://doi.org/10.1103/PhysRevApplied.10.054062}
}

@article{omran2019generation,
  title={Generation and Manipulation of Schr{\"o}dinger Cat States in Rydberg Atom Arrays},
  author={Omran, Ahmed and Levine, Harry and Keesling, Alexander and Semeghini, Giulia and Wang, Tout T and Ebadi, Sepehr and Bernien, Hannes and Zibrov, Alexander S and Pichler, Hannes and Choi, Soonwon and others},
  journal={Science},
  volume={365},
  number={6453},
  pages={570--574},
  year={2019},
  publisher={American Association for the Advancement of Science},
  doi = {https://doi.org/aax9743}
}

@article{heinsoo2018rapid,
  title={Rapid High-Fidelity Multiplexed Readout of Superconducting Qubits},
  author={Heinsoo, Johannes and Andersen, Christian Kraglund and Remm, Ants and Krinner, Sebastian and Walter, Theodore and Salath{\'e}, Yves and Gasparinetti, Simone and Besse, Jean-Claude and Poto{\v{c}}nik, Anton and Wallraff, Andreas and others},
  journal={Phys. Rev. Applied},
  volume={10},
  number={3},
  pages={034040},
  year={2018},
  publisher={APS},
doi = {https://doi.org/10.1103/PhysRevApplied.10.034040}
}

@article{aasen2024readout,
  title={Readout Error Mitigated Quantum State Tomography Tested on Superconducting Qubits},
  author={Aasen, Adrian Skasberg and Di Giovanni, Andras and Rotzinger, Hannes and Ustinov, Alexey V and G{\"a}rttner, Martin},
  journal={Commun. Phys.},
  volume={7},
  number={1},
  pages={301},
  year={2024},
  publisher={Nature Publishing Group UK London},
doi={https://doi.org/10.1038/s42005-024-01790-8}
}

@article{purcell1946resonance,
  title={Resonance Absorption by Nuclear Magnetic Moments in a Solid},
  author={Purcell, Edward M and Torrey, Henry Cutler and Pound, Robert V},
  journal={Phys. Rev.},
  volume={69},
  number={1-2},
  pages={37},
  year={1946},
  doi={https://doi.org/10.1103/PhysRev.69.37},
  publisher={APS}
}

@inproceedings{raimond2006monitoring,
  title={Monitoring the Decoherence of Mesoscopic Quantum Superpositions in a Cavity},
  author={Raimond, Jean-Michel and Haroche, Serge},
  booktitle={Quantum Decoherence: Poincar{\'e} Seminar 2005},
  pages={33--83},
  year={2006},
  doi={10.1007/978-3-7643-7808-0_2},
  organization={Springer}
}

@article{blais2004cavity,
  title={Cavity Quantum Electrodynamics for Superconducting Electrical Circuits: An Architecture for Quantum Computation},
  author={Blais, Alexandre and Huang, Ren-Shou and Wallraff, Andreas and Girvin, Steven M and Schoelkopf, R Jun},
  journal={Phys. Rev. A},
  volume={69},
  number={6},
  pages={062320},
  year={2004},
  doi          = {https://doi.org/10.1103/PhysRevA.69.062320},
  publisher={APS}
}

@article{boixo2018characterizing,
  title={Characterizing Quantum Supremacy in Near-Term Devices},
  author={Boixo, Sergio and Isakov, Sergei V and Smelyanskiy, Vadim N and Babbush, Ryan and Ding, Nan and Jiang, Zhang and Bremner, Michael J and Martinis, John M and Neven, Hartmut},
  journal={Nat. Phys.},
  volume={14},
  number={6},
  pages={595--600},
  year={2018},
  publisher={Nature Publishing Group UK London},
  doi={https://doi.org/10.1038/s41567-018-0124-x}
}

@unpublished{GXC2026,
  author = {Chan, {Guo Xuan} and Lan, Wangwei and Wang,Tenghui and Ma,Xizheng and Deng,Chunqing and Jin,Lijing},
  title  = {System-Level Design of Scalable Fluxonium Quantum Processors with
Double-Transmon Couplers},
  note   = {Manuscript in preparation},
}

\end{document}